\documentstyle[12pt]{article}
\input{epsf}
\newlength{\extraspace}
\newlength{\extraspaces}
\newcommand{\be}{\begin{equation}
\addtolength{\abovedisplayskip}{\extraspaces}
\addtolength{\belowdisplayskip}{\extraspaces}
\addtolength{\abovedisplayshortskip}{\extraspace}
\addtolength{\belowdisplayshortskip}{\extraspace}}
\newcommand{\ee}{\end{equation}}

\newcommand{\bq}{\begin{eqnarray}
\addtolength{\abovedisplayskip}{\extraspaces}
\addtolength{\belowdisplayskip}{\extraspaces}
\addtolength{\abovedisplayshortskip}{\extraspace}
\addtolength{\belowdisplayshortskip}{\extraspace}}
\newcommand{\eq}{\end{eqnarray}}

\newcommand{\newsection}[1]{
\vspace{15mm}
\pagebreak[3]
\addtocounter{section}{1}
\setcounter{equation}{0}
\setcounter{subsection}{0}
\setcounter{footnote}{0}
\begin{flushleft}
{\large\bf \thesection. #1}
\end{flushleft}
\nopagebreak
\medskip
\nopagebreak}

\newlength{\figsize}
\begin{document}
\addtolength{\baselineskip}{.8mm}

\thispagestyle{empty}
\begin{flushright}
{\sc OUPT}-98-64-P\\
cond-matt/9808240 
\end{flushright}
\vspace{.3cm}

\begin{center}
{\large\sc{Scaling Laws for the Market Microstructure of the Interdealer
Broker Markets }}
\\ [12mm]
{\sc  David Eliezer\footnote{deliezer@genre.com}}\\
\vspace{.3cm}
{ Research and Risk Management Dept. \\[2mm]
General Re Financial Products \\
630 Fifth Ave., Ste. 450 \\
New York, NY 10111, USA}\\
\vspace{.3cm}
 and\\
\vspace{.3cm}
 {\sc Ian I. Kogan\footnote{i.kogan1@physics.ox.ac.uk}}\\
\vspace{.3cm}
{Theoretical Physics, Department of Physics \\[2mm]
 University  of Oxford, 1 Keble Road \\[2mm]
 Oxford, OX1 3NP, UK}
\\[12mm]

{\sc Abstract}
\end{center}
\noindent
We discuss a series of simple models for the microstructure of a
double auction market without intermediaries.  We specialize to those
markets, such interdealer broker markets, which are dominated by
professional traders, who trade mainly through limit orders, watch
markets closely, and move their limit order prices frequently.  We
model these markets as a set of buyers and a set of sellers, whose
numbers vary in time, and who diffuse in price space and interact
through an annihilation interaction.  We seek to compute the purely
statistical effects of the presence of large numbers of traders, as
scaling laws on various measures of liquidity, and to this end we
allow our model very few parameters.  We find that the bid-offer
spread scales as $\sqrt{1/{\rm Deal\ Rate}}$.  In addition we
investigate the scaling of other intuitive relationships, such as the
relation between fluctuations of the best bid/offer and the density of
buyers/sellers.  We then study this model and its scaling laws under
the influence of random disturbances to trader drift, trader
volatility, and entrance rate.  We also study possible extensions to
the model, such as the addition of market order traders, and an
interaction that models momentum-type trading.  Finally, we discuss
how detailed simulations may be carried out to study scaling in all of
these settings, and how the models may be tested in actual markets.

\newpage
\newsection{Introduction.}
\renewcommand{\footnotesize}{\small}

Market microstructure has made considerable progress in the past few
years in understanding price formation in formally organized markets.
Most particularly this has been done by considering the dealer's
optimization problem. Starting with the work of Garman
\cite{Garman}, and continuing with Stoll \cite{Stoll}, Ho and Stoll
\cite{HoStoll}, O'Hara and Oldfield \cite{OharaOldfield}, and Amihud
and Mendelson \cite{AmihudMendelson}, the various elements of this
problem and its consequences have gradually been elucidated
\cite{Garman,  Stoll, HoStoll, OharaOldfield, AmihudMendelson,
CMSW81,infoimb,strategy,CMSW78,CHMSW, Demsetz,Tinic,TinicWest72,
TinicWest74,Smidt,SchleefMildenstein,NewtonQuandt,Hamilton78,Hamilton76,
Stoll_Exp,BenstonHagerman,CMSWII,BranchFreed,ORJ}.  For a more complete
bibliography, see \cite{ohara,schwartz}.

A separate but parallel line of research, also beginning with Garman,
and extending through Cohen-Maier-Schwartz-Whitcomb \cite{CMSW81,
CMSW78}, has considered the structure of double auction markets,
markets without intermediaries.  Such markets are simpler to treat
than the dealer problem.  This is because the large number of traders
are considered to act independently of one another and their aggregate
behavior may therefore be treated using well developed stochastic
methods, as a purely statistical system.  Thus it is considered that
the great press of numbers, not any complex social interaction, or
difficult optimization problem, is the principle determinant of large
scale market behavior in a double auction market.  In this regard
Garman \cite{Garman} first studied trade price distributions over time
in a double auction market.  In \cite{CMSW81} Cohen, Maier, Schwartz,
and Whitcomb considered the time to trade in such a market, under the
assumption of finite-sized trader movements.  In \cite{CMSW78}, they
studied the effect of ``thinness'' on market returns, where thinness,
the value of shares outstanding, was used as a proxy for liquidity.

More recently, price evolution in double auction markets was studied
by Bak, Paczuski and Shubik (BPS) \cite{BPS}, who introduced a model
in the same vein as the Garman model, with additional features that
effectively mapped double auction market dynamics onto a type of model
from chemistry, known as diffusion-controlled annihilation.  Like
Garman, they used this to study trade price distributions over time,
obtaining a model that goes beyond the standard Brownian motion model,
to recover the empirically observed ``fat tail'' distributions, and
the associated Hurst exponent.  The diffusion-controlled annihilation
model relevant to markets is a model in one dimensional space, with
initially segregated reactants, and forms an interface, a ``reaction
front''.  This front was first studied by G\'{a}lfi and
R\'{a}cz\cite{gr}, and then later was developed in a series of papers by
Cardy and coworkers\cite{cardy}, and studied numerically by several
groups \cite{d1numerical}.  Early contributors to the theory of
diffusion-controlled annihilation include Doi \cite{doi}, and
Peliti\cite{pel}, and a good review is contained in \cite{MG} ( for a
general introduction into  kinetics of diffusion controlled processes,
see, for example, book \cite{otb}).

In the Garman model traders enter the market at a given price, and if
their trade price is not hit, they leave the market a certain time
later -- they do not change their trade price while they have an open
interest.  This is a behavior characteristic of ``investors'', traders
who buy and sell rarely, and hold for long periods of time, for whom
trading tactics to obtain the optimal trade price are not crucial.
Many markets are in fact dominated by trading of this kind.  However,
there are also markets which are dominated by professional traders,
who typically work for a sell side institution, watch the markets
every minute, and care very much about saving every basis point.  The
best examples this kind of market are the interdealer broker markets,
which exist for swaps, caps, floors, and the treasury market, to name
a few.  This kind of trader places mainly limit orders in the market,
and may change his or her limit order price many times before trading,
based on how he or she perceives the direction of the market, the
direction of related markets, changing hedging needs, and anticipation
of news bulletins.  An external observer might characterize this
behavior as a random walking behavior, and that the collection of all
such traders (i.e. their limit order prices) in this market would then
be seen to diffuse.  This behavior stands in contrast to those studied
empirically by \cite{ORJ}.  We wish in particular to describe the
microstructure of this type of market.  We therefore shall study a
model in which traders both diffuse and annihilate.  We shall then see
that the diffusion property of the model has strong and fascinating
implications for microstructural behavior.

In this paper, like that of Garman and BPS, we shall study a double
auction market as a statistical system.  Rather than studying the
distribution of trade prices, however, we focus on using this model to
calculate scaling laws for expectations of the various quantities that
describe microstructural dynamics, e.g. the bid-offer spread,
time-to-midmarket trade, bid or offer size (or more generally, the
density of traders), and deal rate.  Taken together, these four
quantities describe, somewhat redundantly, the notion of
liquidity. Although liquidity itself has no unique quantitative
definition, each of these quantities is monotonic in liquidity, and we
may define liquidity scaling laws as the functional behavior of each
of these in terms of the others. In this paper we shall consider
liquidity scaling laws as a function of the deal rate.  Existing
intuitions then govern the qualitative behavior of these laws as
follows.

The higher the deal rate, the smaller the bid-offer spread.

The higher the deal rate, the more quickly we can trade at the midmarket.

The higher the deal rate, the greater the density of traders.

The quantities one encounters in studying market microstructure are
measured over short time scales, minutes and seconds.  This feature
simplifies empirical studies, but more importantly, we argue, it
simplifies the nature of the model itself.  This is because the short
time scales leave little or no time for communication between traders.
This implies that we do not need to model the complicated sociological
interactions that take place when large amounts of information passes
between traders -- only the great press of numbers determines scaling
laws for quantities like these.  \footnote{ We distinguish
``sociological'' interactions, the collective interaction of traders
with each other, from ``psychological'' interactions, the response of
individual traders to changes they perceive in the market.  While
sociological interactions ought not be important over these very short
time scales, psychological interactions might, if they were to act in
a correlated fashion.  Our approach carries with it an implicit but
important assertion, that these interactions are in fact uncorrelated
over short time scales, and therefore ``wash out'', so that we may treat
their behavior as a collection of random walks.  }

Making all these intuitions into quantitative scaling laws is the
principal aim of this paper, and these present a set of new and
interesting research problems.  In addition to the scaling laws
associated with various measures of liquidity, there are large sets of
other interesting problems regarding internal market structure over
short time scales.  In this paper, we will also investigate scaling
laws concerned with details of steady state markets, market making
strategies, and a market's response to imbalances between buyers and
sellers.  We note here that scaling laws have been used before to
characterize the behavior of markets -- the best known example is the
Hurst exponent.  However, this example serves to highlight the
difference between our work and previous work, especially \cite{BPS}
-- whereas scaling laws such as the Hurst exponent quantify the
behavior of price evolution in time, and so are scaling laws in the
time variable, ours are concerned with issues of price formation, and
are therefore described by scaling laws with respect to liquidity.

In the next section we will demonstrate how this can be done.  We
shall first introduce a mathematical framework, statistical field
theory, which allows us to treat systems, such as a market, with
variable numbers of random agents, similar in this way to Garman's
model.  Then we will seek a model which is minimal, i.e. the simplest
model possible which describes the market phenomena mentioned above.
We will see how these considerations define this model almost
uniquely, and will find the model to be similar to the dynamics of the
``annihilation'' reaction $B + S \rightarrow \emptyset$ in a
one-dimensional space, a well studied problem (for further reading see
the excellent recent review \cite{MG}).  While Garman \cite{Garman}
was the first to introduce the analogy to this type of reaction into
finance, an important element later added to this picture was the idea
of diffusion.  Bak, Paczuski, and Shubik, \cite{BPS} were the first to
introduce to finance a model with both diffusion and this annihilation
reaction, in a study of price evolution.  Interestingly, it is the
one-dimensionality of the system (a consequence of the standardization
of the trading contract) that gives this model non-trivial dynamics.
In the next section we will present a more complete argument for this
mapping of our problem onto this model, define some financially
relevant variables, and use this to show explicitly how the above
scaling laws, and many more, may be calculated within such a model.

This minimal model is very easily treated, and there already exists a
considerable body of work devoted to it in the physics and chemistry
literature \cite{MG}. When we apply this model to market
microstructure, we shall find that several of the scaling laws and
correlations we seek are calculable by known analytic methods.  This
minimal model is intended to describe market statistics at only the
level appropriate to scaling laws, and to this end we keep the number
of parameters to a minimum.  In addition to the deal rate, we shall
require only two other parameters to define the model completely --
these turn out to be the volatility (or diffusion coefficient) of
individual traders, and the effective width of the price space.  By
calculating and measuring any of the correlation functions these
parameters are easily inferred.

{\bf Going Beyond the Minimal Model}

The notion of studying the scaling laws of a market's internal
structure may be extended beyond markets in the normal state.  In this
paper we shall also introduce models which explore scaling laws in
other settings.  With the exception of the two-fluid model in its
crashing phase, the additional effects of these models are weak, and
modify the dominant effect of statistics only by small corrections.

{\it   --  The Effect of Random Disturbances}

The minimal model with which we begin this study is intended to model
``quiescent'' markets. i.e. those in which deals are steadily ticking
over at a constant rate, and in which no catastrophes are occurring.
We may extend ourselves beyond this most ideal regime into others, by
first considering the effects of random disturbances within the
market.  We can do this by imagining that they cause trader dynamics,
defined by their drift, volatility or flux, to jump around randomly.
The strength of this random noise component of these quantities
behaves somewhat like a temperature, i.e. the amount of ``energy''
injected into the market by the influence randomly occuring series of
external events.  We imagine that the market is continually buffeted
by a series of small jolts, that force it partially into its excited
states.  The strength of this noise controls the weighting of the
system in these states, and introduces noise-strength dependent
corrections to the correlations of market observables.

{\it   -- Scaling Laws in the Crashing Phase: the Two Fluid Model}

A market which is in the process of a long crash does not look
quiescent.  However, it is possible to regard a market in free-fall as
a market in a different phase, subject to a different sort of
dynamics, but one which may be probed by the same sort of statistical
scaling laws nevertheless.  It is not sufficient, for this purpose, to
simply consider the minimal model in the presence of a large imbalance
between buyers and sellers, because this would not recover the well
known phenomenon of the widening of the bid-offer spread.  We believe
that the missing element, ignored by the minimal model, is market
order traders.  The minimal model contains only traders who enter
limit orders into the system, so that the true best bid/offer can be
seen on screen, but this would not represent the reality of a crash,
during which most traders attempt to hit the best bid directly, as a
market order.  To address this, we introduce a ``Two-Fluid Model'',
one in which there are two flavors of buyer, $B$ (limit order
bidders)and $B'$ (market order bidders), and two flavors of seller $S$
(limit order sellers) and $S'$ (market order sellers).  The prices of
the unprimed traders {\it only} are used to calculate the best bid and
offer on the trade screen, and are therefore visible to other traders.
Thus the Two-Fluid Model includes the reactions $B + S \longrightarrow
0$, $B' + S \longrightarrow 0$, and $B + S' \longrightarrow 0$, but
not $B' + S' \longrightarrow 0$, because these primed traders are
invisible to each other.  This model, when the population of invisible
$S'$ traders is large, now has the familiar behavior in which the
bid-offer spread widens out during a crash.  We pause here to note
that Bak, Paczuski, and Shubik introduced ``heterogeneous'' models,
too, ie. models involving more than one types of trader \cite{BPS}.

{\it   -- The onset of crashes: the Bias Model}

We have argued that the short time scales of our measurement enable us to
ignore collective, sociological effects of trader behavior, because
they will not have had time to communicate appreciably.  But, if we were
to consider slightly longer time scales, it would be easy to identify the 
most
direct and likely medium for this communication -- it is the trading
screen itself, which instantly informs traders of trade prices.  Over
slightly longer time scales, it also allows them to observe trends.  We 
may get
more ambitious and attempt to go beyond the very shortest time scales, to
model the most immediate of collective behavior.  We treat an interesting
model of this type in this paper, as well.

We introduce a simple model that treats trend-following behavior in
traders under the name of the ``Bias Model''.  The rate at which
trend-following or ``bias'' builds up in this model is controlled by an
external parameter, which we call the ``market tension'', a property of the
market which we will measure, and which we expect to be slowly varying.
When this trend-following parameter is positive, the traders' drift is
increased each time the trade price increases, and decreased each time
the trade price decreases -- this is meant to correspond to
momentum-trading.  When bias is negative, so that the traders' drift
moves in the direction opposite to the trade price, this corresponds to
bargain-hunting/profit-taking.

With positive tension, this model is intended to develop instabilities 
which
mimic market crashes, and preliminary simulations confirm that this is
so. The model builds up bias slowly at first and, depending on the market
tension parameter and the starting point of the trader drift, the market
eventually builds up enough drift to leave the diffusion-dominated
regime.  At this point it goes into a ballistic regime, in which drift
plays a significant or dominant role -- the crashing phase.
We propose to use this model to study scaling laws in market
microstructure in the prelude to a crash.

{\it   -- The Diffusion Proposal and the Dealer's Optimization
Problem: A modification of the order arrival process}

Although it is outside the main development, we shall also use our
dynamical framework to consider a problem in market making.  We ask
the question: what is the optimal bid-offer spread that a monopolist
market maker can set, as a function of time, given that he has a view
of all of the limit orders placed in the market?  This dealer
optimization problem has been considered in great detail by Garman
\cite{Garman}, Stoll\cite{Stoll},Ho and Stoll\cite{HoStoll}, O'Hara
and Oldfield \cite{OharaOldfield}, among many others.  We do not wish
in this paper to engage this much more complicated topic -- our intent
is simply to suggest that the Poisson process model for order arrival may 
be
refined by using the known initial data of limit orders and the
diffusion proposal, to obtain a sharper estimate of order arrival.  We
introduce a simple model for maximizing profit while minimizing the
net open position, which illustrates this method.

{\bf Testing the Models}

In the last section of this paper we shall discuss the testing of the
predictions of these models.  We have consciously kept the number of
parameters describing the model to a minimum, so as to isolate the
effect of large numbers of traders and to make the model easily
testable.  Our testing will be done in a future publication, and will
concern markets with institutional traders, and measurements will be
made over very short time scales (minutes and seconds), in a single
market \cite{ftrexptl}.  In this section we review the previous
studies of the spread (principally in dealer markets, which dominated
by the long term investors, not the short term traders we are
interested in).  We also lay out the program for testing
experimentally the scalings we have computed, and will compute, for
these models.

{\bf A Brief History of Physical Approaches to the Financial
Markets}

The application to financial markets of analogies to physical systems
have a long history starting from Louis Bachelier\cite{bach}, who
first proposed the random-walk model of the stock market.  In
the 1960s and 1970s these ideas became very popular and eventually lead to
the famous Black-Sholes option pricing formula \cite{BS} and to
Mandelbrot's applications of scaling behaviour to financial markets
\cite{mandelbrot}.  Recently, a lot of papers have appeared in which
markets were treated as far-from-equilibrium dynamical systems
\cite{AAP}.  It is impossible here to review all relevant recent
papers and areas of research of the quickly growing field of
econophysics \cite{contarchive}.  However, we pause here to mention
several notable examples, such as work on scaling behaviour for
exchange rates \cite{scalingexchange}, ``log-periodic'' oscillations
as crash precursors \cite{log}, dynamics of the interest rate curve
\cite{interestrate}, and market fluctuations \cite{market,BPS}.  There
are, of course many other deserving papers which we do not have the
space to include.

\newsection{Development of the minimal model}

In this section, we present an argument for the mapping of
microstructure of the interdealer broker markets onto a statistical
field theory, in which particles diffuse and annihilate, a well known
model of chemical reactions.

{\it Configuration Space of the Model}

What is the statistical system suggested by our arguments above?  Because
we seek to compute only the most dominant behavior as a function of
liquidity, without the fine detail, we construct first of all the
simplest possible model which is consistent with the proposal described
in the introduction.

\begin{figure}
\begin{center}
%\leavevmode
\epsfxsize = 0.95\figsize
\epsfysize = 0.25\figsize
\epsffile{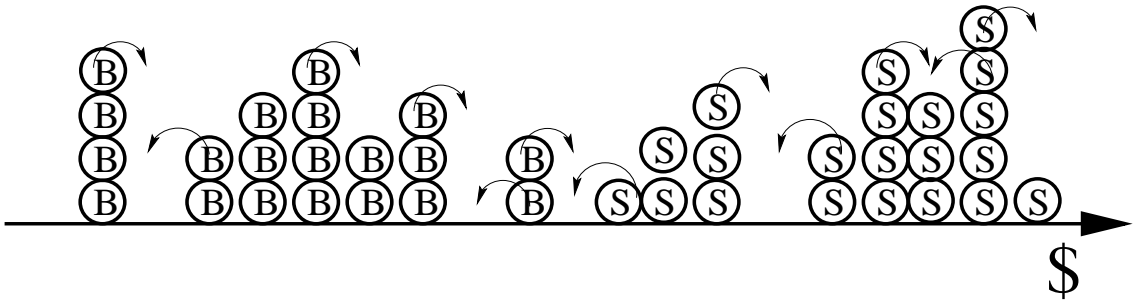}
\end{center}
\caption{\label{1}
A typical  distribution of buyers (balls
 with letter B) and sellers (balls with letter S). Arrows describe
 ``hops'' - not all buyers and sellers will hop in each instant of time.}
\end{figure}

Within a simple market, we wish to investigate the scaling of the {\it
market observables} i.e. the best bid, best offer, trade price, deal
rate and other things we might infer from a trading screen.  In
particular, we wish to study their dependence on the number of traders
(buyers and sellers) present in the market.  This requires us to treat
the dynamics of the market observables as functions of the underlying
dynamics of {\it individual} traders.  We do this as follows.  To each
buyer we associate a price, his bid price, and to each seller his
offered price, and imagine the two types of trade prices moving around
on a 1 dimensional space we call the ``price space'' ${\cal P}$.  (See
Figure \ref{1}) (Of course, we start our model off by placing all seller
prices above all buyer prices. )  It is convenient to regard this
space as a discrete one-dimensional lattice on which traders
(i.e. their trade prices) ``hop'' from site to site.  In this model
each trader trades in a single standard size \footnote{ This is not a
terribly strong restriction, because interdealer broker markets often
have a standard size in which much of their trading is done, and are
often willing to break up larger trades if necessary}.

In this setting, each of our market observables is calculable.  We shall
compute a statistical average of the market observables, and this
average will run over the possible positions of all the traders, over
a suitable probability measure, specified at a time $t$.  This
probability measure we refer to as the ``state'' of our system.

{\it Dynamics of the Model}

\begin{figure}
\begin{center}
%\leavevmode
\epsfxsize = 0.95\figsize
\epsfysize = 0.25\figsize
\epsffile{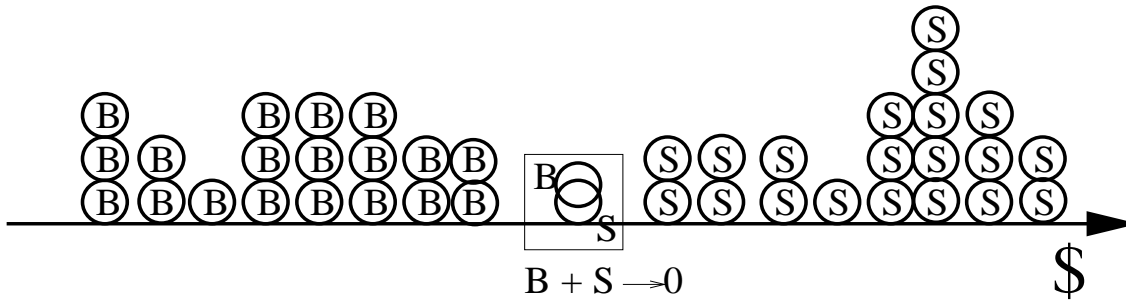}
\end{center}
\caption{\label{2}
Annihilation  $B + S \rightarrow 0$}
\end{figure}

\begin{figure}
\begin{center}
%\leavevmode
\epsfxsize = 0.95\figsize
\epsfysize = 0.25\figsize
\epsffile{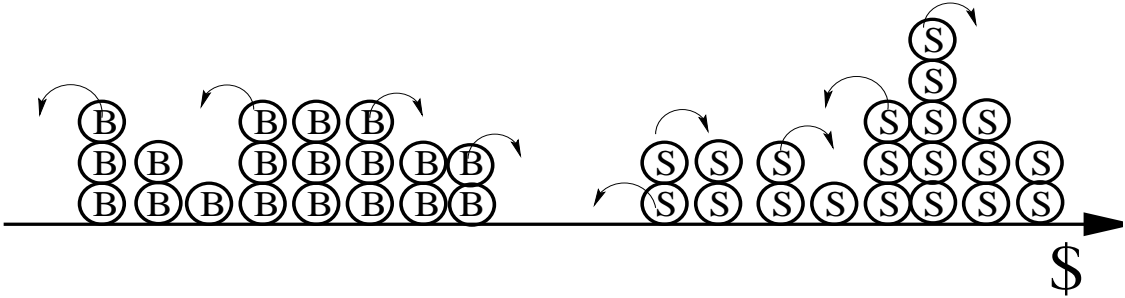}
\end{center}
\caption{\label{3} A new distribution of buyers (balls with letter B)
and sellers (balls with letter S) after an annihilation (trade) has
occured. Note that the bid-offer spread fluctuates. Here it is bigger
than in Figure \ref{1} }.
\end{figure}

We wish to study not only the static behavior of the best bid or
offer, but the dynamics as well, so we must specify a dynamic law for
the movements of the traders.  This law is not completely arbitrary.
Any dynamic law we may specify must guarantee that when a buyer and a
seller are at the same point in price space, they do a deal and
disappear from the market.  At each moment in time, therefore, the
dynamical law deletes from the state any pair of buyers and sellers it
finds occupying the same point in price space.  We call this part of
the dynamics ``annihilation'' (See Figures \ref{2} and \ref{3}).

As discussed in the introduction, an essential ingredient of our model
is that traders in the interdealer markets change their trade prices
often while they have an open interest, in an effectively random,
uncorrelated fashion and this leads us generally toward a random walk
behavior for individual traders, and diffusion for the behavior of the
aggregate.  The {\it precise} nature of this random behavior in price
space is somewhat hard to know, and depends on a trader's psychology,
etc.  We do not wish to attempt to treat the details of these effects,
only the effect of the weight of numbers of traders present, an effect
which any more detailed theory would also contain.  Our model in this
sense should be considered ``minimal''.  In keeping with the modest
goal of capturing the gross, purely statistical features of
random-walking traders only, we introduce the simplest random walk
behavior, with zero drift and constant volatility for the traders
(hereinafter known as $D$, the trader diffusion constant, or trader
volatility). Hence, our minimal model for a trader's price evolution
process will consist of diffusion interspersed with annihilation,
i.e. it is described by a {\it diffusion controlled (or diffusion
driven) annihilation reaction}.  In the physics and chemistry
literature this sort of dynamics has been studied extensively, and a
good introduction is contained in \cite{otb}, and references therein,
as well as \cite{MG}.  In addition, the mapping of market dynamics
onto diffusion-controlled annihilation was first used by \cite{BPS} to
study the random process governing trade price evolution.

{\it Distribution of Traders}

To compute the scaling laws introduced in the previous section, we must
impose a probability measure, the state measure, on the positions of each
type of trader.  Although we have specified a dynamical law for the state
measure, we haven't yet defined it until we have also specified its
initial conditions.  This represents a lot of missing information, which
cannot be deduced from the conditions we have so far imposed on our model
-- we must discover what kind of diffusion-annihilation dynamics is
relevant to our problem.  We do this as follows.  It is natural, for this
kind of problem, to restrict our attention to quiescent markets,
i.e. those where catastrophes such as market crashes or large, explosive
jumps due to the injection of critical new information are {\it not}
happening \footnote{Because our time scales are so short, there should
be many such intervals during the day in most markets.  Indeed, most of
the time, by this definition, most markets are ``quiescent''. }.  We
assert here that a quiescent market is an {\it approximately steady
state} market, and thus we define our evolving state measure, at any time
$t$, to be that probability measure which is the steady state solution of
our diffusion-annihilation dynamics.

As a first approximation, we allow traders to enter the market only at
the ends, as though buyers began by bidding very low, and sellers by
offering very high, and then improving.  This is clearly an
approximation which we expect will not change the dynamics too
greatly.  In a series of numerical simulations we will address the
question of more realistic insertion processes, in a future 
publication\cite{ftrnumerical}.

Like most such models we may solve this minimal model by simulation.
However, we can go much further for this model, because it happens to
be similar to many well-known analytically soluble
models\cite{BetheAnsatz}, see also \cite{MG} and may itself be soluble
-- in addition, there are approximate analytical methods available
\cite{cardy}.  Through these, the set of steady state solutions is
well understood, and it is interesting and helpful to note that the
set of steady state solutions corresponds to the lowest energy states
(``ground states'') of the equivalent physical system, and is a family
indexed by a single parameter, the deal rate $J$.  Thus, given a deal
rate, i.e. a market at a given level of liquidity, there is a unique
steady state solution to our model.  The scaling laws in question may
be evaluated simply as an expectation value of the market observable
in any such ground state, regarded as a function of the choice of the
ground state.  Making use of this, many numerical results for scalings
have already been found, see for example \cite{d1numerical}.

{\bf The dynamical framework: statistical field theory}

{\it The State is an Evolving Measure on Sequences on the Price Space}

In this section, we introduce the mathematics of our dynamical
framework.  It is a variant of statistical field theory, first used
 by Doi \cite{doi} and Peliti \cite{pel}, and it is similar to the
approach taken by Garman in his double auction model \cite{Garman} and
Bak-Paczuski-Shubik\cite{BPS}.
The scaling laws we seek are moments of the distribution of buyers and
sellers at a slice of time $t$.  Therefore, we do not need the full
distribution on path space, but just a time-slice of this
distribution.  In summary, we require a distribution on the positions of
traders at a time $t$, which evolves in time according to
diffusion-annihilation dynamics.  It is this distribution, hereafter
known as the ``state measure'' or just the ``state'', which shall be
the dynamical variable of our model.

The measure space underlying this system is different from those one
usually encounters in financial problems.  Stochastic problems in finance
are usually expressed as stochastic differential equations, which lead to
a distribution function on an underlying sample space, say ${\cal R}^N$, of
fixed dimension.

If $N$ traders obeyed a simple diffusion law, instead of
diffusion-annihilation dynamics, we would describe the system with a set
of coupled stochastic differential equations.  Solving this system would
give us an evolving state measure, $\mu \in {\cal M}_N$, a measure on the
$N$-dimensional underlying sample space (${\cal R}^N$, in this case).  In 
our
problem, however, the dimension of the underlying sample space itself can
change during the evolution, because traders may exit or enter the
system.  The measure space in which there are exactly $N_T$ traders
(i.e. ``fixed trader number'') is but a small part of the full measure
space of our system.  In fact, we would like our model to allow the
possibility that the system may, along one branch, have had a buyer and
seller collide, annihilate, and remove each other from the system, and
along the other branch that they did not meet, and therefore still remain
within our system, in other words, that the system resides {\it partly}
in one trader number subspace ${\cal M}_{n_b,n_s}$ and partly in another,
${\cal M}_{n_b-1,n_s-1}$.  Here let ${\cal P}_{n_b,n_s}$ be an $n_b\cdot
n_s$ dimensional underlying sample space in which $n_b$ buyers and $n_s$
sellers are present -- then ${\cal P}_{n_b,n_s}$ is a small part of the
full underlying sample space ${\cal P}_{\rm Full} = \cup_{n_b,n_s} {\cal
P}_{n_b,n_s}$.  ${\cal M}_{n_b,n_s}$ is the space of measures on ${\cal
P}_{n_b,n_s}$, and ${\cal M}_{\rm Full}$ is the space of measures on
${\cal P}_{\rm Full}$.  Then we are required to include in our full
measure space linear combinations of distributions on all the different
underlying sample spaces ${\cal P}_{n_b,n_s}$: $\mu =
\sum_{n_b,n_s} c_{n_b,n_s} \mu_{n_b,n_s}, \mu \in {\cal M}, \mu_{n_b,
n_s} \in {\cal M}_{n_b, n_s}$ (where ${\cal M}_{n_b,n_s}$ is the space
of measures on the underlying sample space ${\cal P}_{n_b,n_s}$), and
therefore the full measure space for our stochastic system ${\cal
M}_{\rm Full}$ is a vector space direct sum of the measure spaces with
fixed trader number ${\cal M}_{\rm Full} = {\oplus}_{n_b,n_s} {\cal
M}_{n_b,n_s}$.  ${\cal M}_{\rm Full}$ is the space of measures on the
full underlying sample space ${\cal P}_{\rm Full}$.  The norm of the
projection into each subspace ${\cal M}_{n_b,n_s}$ is the probability
of having $n_b$ buyers and $n_s$ sellers. As these probabilities must
add, the norm on ${\cal M}$ is the sum of the fixed subspace norms
$||\mu|| = \sum_{n_b,n_s} c_{n_b,n_s} ||\mu_{n_b,n_s}||, \mu \in {\cal
M}, \mu_{n_b, n_s} \in {\cal M}_{n_b, n_s}$.

This dynamical system differs in one further respect from systems we
would describe with stochastic differential equations.  The identity of
traders does not matter to us -- if buyer 1 sits at site 1 and buyer 2
sits at site 2, that is the same, for us, as if buyer 1 sat at site 2 and
buyer sat at site 1.  As it stands, the system keeps track of more
information than we are concerned with.  We therefore identify these
different configurations as the same point in our underlying space, by
modding out by the action of the permutation groups (separately for
buyers and sellers) on that space, i.e. ${\cal P}_{n_b,n_s} = (({\cal
P})^{n_b}/S^{n_b}) \otimes (({\cal P})^{n_s}/S^{n_s})$.  Modding out
${\cal P}^n$ by $S^n$ is the same as restricting the measure space on
${\cal P}_n$ to symmetric measures on ${\cal P}^n$ and, to compensate
for multiple counting of configurations, adjusting the measure by an
appropriate multinomial factor \footnote{This factor is not a constant on
${\cal P}^n$, and therefore this multinomial factor is in fact an
operator, which we describe in the following subsections.  }.

A point in ${\cal P}_{n_b,n_s} = (({\cal P})^{n_b}/S^{n_b}) \otimes
(({\cal P})^{n_s}/S^{n_s})$ quotient space defines only the number of
buyers and sellers sitting at each point in the price space ${\cal P}$
(the ``occupation numbers'' $n^b_x$ and $n^s_x$ at $x \in {\cal P}$), and
is thus defined by two summable sequences of positive integers on ${\cal
P}$, $\{ n_b(x), x \in {\cal P}\} \in {\cal L}_1({\cal P, \cal Z^+})$
(for the buyers) and $\{ n_s(x), x \in {\cal P}\} \in {\cal L}_1({\cal P,
\cal Z^+})$ (for the sellers).  Here ${\cal L}_1({\cal P, \cal Z^+})$ is
the space of positive-integer valued, summable sequences on ${\cal P}$.
In other words, there is a bijection between ${\cal L}_1({\cal P, \cal
Z^+}) \otimes {\cal L}_1({\cal P, \cal Z^+})$ and our full space of
configurations ${\cal P}_{\rm Full} = \cup_{n_b,n_s} {\cal P}_{n_b,n_s}$.
This in turn defines a norm-preserving bijection between ${\cal M}({\cal
L}_1({\cal P, \cal Z^+}) \otimes {\cal L}_1({\cal P, \cal Z^+}))$ and
${\cal M}( {\cal P}_{\rm Full} )$.  This aspect of statistical field
theory is much like the Garman model\cite{Garman}.  Thus our states are
elements of the measure space ${\cal M}( {\cal P}_{\rm Full} ) = {\cal
M}({\cal L}_1({\cal P, \cal Z^+}) \otimes {\cal L}_1({\cal P, \cal
Z^+}))$.

{\it The Pure State Basis for the Space of State Measures, and Insertion
and Deletion Operators}

The foregoing describes the mathematical system required by our
proposal to study liquidity scaling laws within the simplest possible
market model.  It turns out that the configuration space we have
deduced above is identical to that of statistical field theory, used
in physics and chemistry to describe systems with a variable number of
particles.  Below we include a brief qualitative introduction to
statistical field theory.  We have modified it slightly, following the
work of Doi\cite{doi} and Peliti\cite{pel}, in order to remove some
specifically quantum mechanical features which appear in standard
treatments of the field\cite{iz}.

Statistical field theory is most easily introduced in a particular
basis for the measure space, that of the ``pure states''.  A pure
state is an element of ${\cal M}( {\cal P}_{\rm Full} ) = {\cal
M}({\cal L}_1({\cal P, \cal Z^+}) \otimes {\cal L}_1({\cal P, \cal Z^+}))$
which has all of its probability mass concentrated at a single point
of ${\cal P}_{\rm Full} = {\cal L}_1({\cal P, \cal Z^+}) \otimes {\cal
L}_1({\cal P, \cal Z^+})$.  We may describe such a point with the
sequences of occupation numbers $\{n^b_x \}$, $\{n^s_x\}$ -- denote
this state as $|\{n^b_x, n^s_x: x \in {\cal P}\}>$.  These states form
a natural basis for ${\cal M}({\cal L}_1({\cal P, \cal Z^+}) \otimes
{\cal L}_1({\cal P, \cal Z^+}))$.  The usual ${\cal L}^1$ norm of a
measure $\mu = \sum_{\{n_x^b,n_x^s\}} P_{\{n_x^b,n_x^s\}} | \{
n_x^b,n_x^s\}>$ in this space is simply $||\mu|| =
\sum_{\{n_x^b,n_x^s\}} P_{\{n_x^b,n_x^s\}}$.

Formally, this basis is constructed as follows, in the example of a
discrete price space such as ours.  Assume, to start with, that there
is only one type of trader, say buyers, present in the market.  We
define the unique state $|0>$ which is empty of traders, and linear
operators ${\Psi^{(B)}_x}^\dagger, x \in {\cal P}$ which insert a
buyer at $x$, ${\Psi^{(B)}_y}^\dagger |\{n_x : x \in {\cal P}\}> =
|\{n_x : x \in {\cal P}, x \ne y, {\rm and\ } n_x + 1, x=y\}>$.  Then
an element of the basis alluded to above is \bq |\{n_x : x \in {\cal
P}\}> =\prod_{x \in {\cal P}} ({\Psi^{(B)}_x}^\dagger)^{n_x}|0>.  \eq
These operators are known as buyer insertion, or buyer creation
operators.  In a similar way we introduce the buyer deletion, or buyer
destruction operators $\Psi^{(B)}_x, x \in {\cal P}$, $\Psi^{(B)}_y
|\{n_x : x \in {\cal P}\}> ~~ = |\{n_x : x \in {\cal P}, x \ne y,
{\rm and } n_y - 1, x=y\}>$ if $n_y > 0$, and $0$ otherwise.  The
linear operators $\Psi^{(B)}_x$ and ${\Psi^{(B)}_y}^\dagger$ commute
with each other unless $x = y$, in which case we have
${\Psi^{(B)}_x}^\dagger\Psi^{(B)}_x -
\Psi^{(B)}_x{\Psi^{(B)}_x}^\dagger = 1$, in other words
${\Psi^{(B)}_x}^\dagger\Psi^{(B)}_y -
\Psi^{(B)}_y{\Psi^{(B)}_x}^\dagger = \delta_{xy}$, where $\delta_{xy}$
is the Kronecker delta. (In the case where ${\cal P}$ is a continuous
space, we write the above as ${\Psi^{(B)}_x}^\dagger\Psi^{(B)}_y -
\Psi^{(B)}_y{\Psi^{(B)}_x}^\dagger = \delta(x-y)$, where the right
hand side is Dirac's delta function.)  For each $x \in {\cal P}$ we
may define operators $N^{(B)}_x$ whose eigenvectors are the basis
elements $|\{n_x : x \in {\cal P}\}>$, and whose eigenvalues are the
occupation numbers $n^b_x$ as $N^{(B)}_x = {\Psi^{(B)}_x}^\dagger
\Psi^{(B)}_x$.  This operator goes by the name of the occupation
number operator, but in our context it is sensible to call it the
buyer number operator.

{\it The Norm on our Measure Space -- a Single Kind of Trader}

With this in hand, we may define the appropriate statistical field
theory norm $<\cdot>$ (due to Doi\cite{doi} and Peliti \cite{pel}) in
terms of the function ${\cal F}$ on pure states ${\cal F}(\mu) = 1$ if
$\mu = | 0 >$ and $0$ otherwise, which we extend as a linear operator
to all states.  In terms of this function ${\cal F}(\cdot)$, we define
$< \mu > = {\cal F}( \exp( \sum_x \Psi^{(B)}_x ) \mu )$.  Here the
exponential factor is precisely the operator that introduces the
correct multinomial factors into the subspace measures, to compensate
for overcounting of configurations in our representation of the
configuration space as positive integer sequences on ${\cal P}$.
Because of the presence of this factor, the averaged density of
traders may be calculated as $<{\Psi^{(B)}}^\dagger(x)\Psi^{(B)}(x)> \
\ = \ \ <\Psi^{(B)}(x)>$.  Note that the coefficients $P_{\{n^b_x\}}$
of the pure states are proportional to, but not equal to, the
probability of the occurrence of that state.  We refer to this as the
``probability amplitude''.

This treatment of Doi's work is different from that in his paper, so
here we relate our exposition of Doi's work to his own.  Statistical
field theory is normally used in the context of quantum theory, which
demands a conjugation operation, an ${\cal L}^2$ norm, and inner
product, making the space a Hilbert space ${\cal H}$: $a, b \in {\cal
H} \longrightarrow <a|b> \in {\cal C}$.  Doi embeds his formulation
inside this Hilbert space.  He does this by introducing a special
state $<SUM|$
\bq
<SUM| = <0|\prod_{i} \exp(\sum_x \Psi(x))
\eq
with the property
$$
<SUM|\Psi^\dagger = <SUM|
 $$
and he defines the ${\cal L}^1$ norm in terms of the ${\cal L}^2$
space inner product as $||\Psi||_1 = <SUM| \Psi>$

{\it The Norm on a Space with a Second Type of Trader: Sellers}

We may of course have more than one type of trader in our system, and it
is essential that we do.  We introduce a basis containing a complete set
of occupation states for sellers, so that a full basis for our measure
space is $|\{n^b_x, n^s_x : x \in {\cal P}\}>$, and along with the buyer
insertion and deletion operators ${\Psi^{(B)}_x}^\dagger, \Psi^{(B)}_x$, we
introduce the seller insertion and deletion operators
${\Psi^{(S)}_x}^\dagger$, $\Psi^{(S)}_x$.  Similarly we define the seller
number operator $N^{(S)}_x = {\Psi^{(S)}_x}^\dagger \Psi^{(S)}_x$.
The standard norm of a measure $\mu = \sum_{\{n_x^b\}, \{n_x^s\}}
P_{\{n_x^b\}, \{n_x^s\}} |\{ n_x^b\}, \{ n_x^s \}>$ in this expanded space 
is
simply $\sum_{\{n_x^b\}, \{n_x^s\}} P_{\{n_x^b\}, \{n_x^s\}}$, and so
the norm we will use for expectations is $<\mu> = {\cal F}( \exp( \sum_x
\Psi^{(B)}_x ) \exp( \sum_x \Psi^{(S)}_x ) \mu )$

{\it We May Allow Trader Types with Several Different Kinds of Behavior}

In analogy with the case of sellers, we may introduce completely new
species of traders, such as the market order traders of the two-fluid
model, or traders with different dynamical behavior (e.g. momentum
traders, as in the bias model, see section 4).  We may even introduce
other species still, which are not traders at all, but serve to
communicate information between traders (see appendix).  A species may
have a completely different kind of dynamics, or even no dynamics at all.
One important example of additional species that we shall need to
introduce is the non-diffusing, externally controlled trader (either
buyer or seller) which annihilates traders of the other type, but does
not diffuse, and may not leave the market after trading (e.g. a
specialist who sets a bid price and an offered price, good for any size,
and stays in the market (see section 3).  .  We denote these as
$\Psi^{(\cdot),(e)}_x$, ${\Psi^{(\cdot),(e)}_x}^\dagger$, where $(\cdot)$
stands for either $(B)$ or $(S)$.

{\it The Fundamental Variables of the Theory are Operators on the
State Measure}

The linear operators $\Psi^{(\cdot)}_x$, ${\Psi^{(\cdot)}_x}^\dagger$,
and $N^{(\cdot)}_x$ are actually maps from the price space ${\cal
P}$ into the set of linear operators on the measure space.  Following
the terminology of statistical field theory, we shall refer to such
operators as ``field operators'' or ``fields''.

The $\Psi$\,s and $\Psi^\dagger$\,s are the building blocks of all
computation in statistical field theory.  Consider any market observable,
e.g. the best bid $B$.  It is possible to define (in terms of $\Psi$\,s and
$\Psi^\dagger$\,s) an associated linear operator $\hat B$, that acts on
pure states $|{n_x}>$ by multiplication by the best bid of that pure
state.  The action of this operator on the measure defining the state of
our system $|{\rm state}> = \sum_{\{ n_x \}} P_{\{ n_x\}} |{\{n_x\}}>$
gives another measure $|b> = \hat B |{\rm state}>$, whose expansion
coefficients in the pure state basis is the best bid of the pure state
times its probability amplitude in $|{\rm state}>$:  $|b> = B_{\{n_x\}}
P_{\{n_x\}} |\{ n_x \}>$.  Its norm is therefore equal to the expectation
value of the best bid $<B>_t = ||\hat B |\rm state, t> ||$

We may use this procedure for evaluating any desired expectation value in
statistical field theory.  We construct the appropriate operator using
the $\Psi$s and $\Psi^\dagger$s, apply this operator to our (time
dependent) state measure $|{\rm state},t>$, and evaluate its norm,
thus $<{\rm Op}> = || \hat{\rm O}{\rm p} |{\rm state}, t> || $.
We construct several of these operators  in the next section.

We shall use this procedure throughout this paper, but in a modified
form.  The time evolution operator $U_{t,t'}$ satisfies $|{\rm state},t'>
= U_{t,t'} |{\rm state},t>$.  Thus, the expectation of any observable $A$
at a time $t$ is$<A>_t = || \hat A |{\rm state},t> || = || \hat A 
U_{t,0}|{\rm
state},0> || = || U^{-1}_{t,0}\hat A U_{t,0}|{\rm state},0> ||$.  We
define the time dependent version of an operator $A$ via its conjugation
with $U$: $\hat A(t) = U^{-1}_{t,0} \hat A U_{t,0}$.  We may write
operator expectations as $<A>_t = || \hat A(t) |{\rm state},0> || $.

{\it Dimensional Analysis and the Parameters of Diffusion-Controlled 
Annihilation}

Diffusion-annihilation dynamics consists of diffusions interspersed with
annihilations, for each species of trader.  We may construct both a
diffusion operator $\cal{D}_{(\cdot)}$ and an annihilation operator
$\cal{U}_{\rm ann}$ from the set $\{\Psi^{(\cdot)}_x,
{\Psi^{(\cdot)}_x}^\dagger\}, x \in {\cal P}$, in order to assemble a
diffusion-annihilation operator that evolves a state forward in time
within our model.  Although we shall postpone a discussion of these
operators to the appendix, we shall make one important comment about the
dynamics of the model.  The operator evolution equation ( see appendix
\ref{evform}) may be differentiated to yield an operator differential
equation.  In general, if a partial differential equation is invariant
under the change of dependent variables $x \longrightarrow \alpha x$, $t
\longrightarrow \beta t$, then its solutions are also invariant, and this
is a powerful constraint on the form of its solutions.  More commonly,
however, PDEs are changed into a different, but similar equations.
However, this invariance property may be given to any PDE at all,
provided that there are constants in front of each term that are defined
to scale in the appropriate way under the abovementioned transformation,
and by this method we may infer strong constraints on the way the
solution depends on these constants, those in the boundary conditions, and
the dependent variables $t$ and $x$.  The scaling of these constants is
known as their dimensions, and the (generally very simple) analysis to
determine the constraints they imply on the solutions is known as
dimensional analysis.

In our case, there is only one coefficient necessary in the PDE, the
diffusion coefficient $D$, with dimensions $x^2/t$, i.e. $D \sim ({\rm
dollars})^2/({\rm sec})$.  In addition, there is a dimensionful constant
in the boundary conditions, $J$, the rate at which traders enter the
market at the boundaries, which in a steady state market is equal to the
deal rate, with dimension $J \sim 1/({\rm sec})$.  Thus the expectation
$<X>$ of any quantity $X$ with dimensions $[X] = (dollar)^m /sec^n$ must
be proportional to $ (D/J)^{m/2} J^n \sim J^{n-m/2}$.  In particular,
lengths in the model scale as $\sqrt{D/J}$.  As the foregoing analysis
has been for a continuous price space, note that the quantities $<X>$
must converge to something well-defined in the continuum limit.

In the foregoing, we have defined the number operator $N^{(\cdot)}(x)$,
an operator which gives the number of traders sitting at point $x$, in a
pure state.  It is, of course, dimensionless.  This operator is very
closely related to another operator, the density operator
$\rho^{(\cdot)}(x)$, which gives the number of traders per unit length
near point $x$.  They are related as $\rho^{(\cdot)}(x) =
N^{(\cdot)}(x)/\delta S$, where $\delta S$ is the lattice spacing.  In 
this case it
is the trader density $<\rho(x)>$, and not $<N(x)>$, which has a good
continuum limit, and so $<\rho> \sim \sqrt{J/D}$

Note that our discussion of dimensional analysis has ignored the lattice
spacing $\delta S$, which is actually another dimensionful quantity on 
which
these scalings may depend.  The length of the price space $L$ has also
been omitted.  These two represent the lower and upper limits of the
length scales in our model, and if we were to imagine a Fourier analysis
of the dynamics, most of the dynamics takes place far from these two
scales, in the intermediate regime.  As a result, dynamical quantities
tend to depend slowly at best on these quantities, often adding
logarithmic corrections to scaling laws.  As a result, it is correct,
generally speaking, to ignore these length scales in a dimensional
analysis of scaling behavior.  Numerical simulations of this model
\cite{d1numerical} have borne out these arguments.

{\it Parametrizations of the Reaction Front}

A model of this kind allows us to calculate the relationships described
in the introduction, but also many more.  In general, if we start this
model with any initial state satisfying the initial condition above, it
will evolve to a state with a ``reaction front''.  This is a region
within the state where the buyers meet the sellers.  At the center of the
reaction front is, of course, the best bid, best offer and midmarket.
Beyond the best bid/best offer we expect to see an increasing density of
buyers/sellers.

A stationary reaction front corresponds to an approximately steady state
situation similar to the ground state introduced in the previous section.
Buyers and sellers approach each other, meet near the center, and
annihilate.  At the same time, traders
are injected into the system at a rate equal to the deal rate.

The statistics that we read off of a trading screen describe the shape
and dynamics of the reaction front -- the best bid locates the top of the
lower edge and the best offer the bottom of the upper edge (see
Figures 1 and 3).
The bid size tells us the height at the best bid point, and the offer
size tells us a similar thing about the best offer point.  And the last
trade field tells us about its recent history.

Generally we are interested in all parameters which describe this
reaction front, its shape and its dynamics.  The bid-offer spread is
but one, albeit the most important, of the parameters describing the
shape of the reaction front.

Within a pure state, we may describe the shape fully, at a time $t$, by
specifying the number of buyers $\rho_B(x,t)$ and sellers $\rho_S(x,t)$
at each point $x$ of the price space.  Obviously $\rho_B(x,t)=0$ for $x$
greater than the best bid, and $\rho_S(x,t)=0$ for $x$ less that the best
offer.  The number of sellers between $x_\ell$ and $x_u$ is then
$\sum_{x=x_\ell}^{x_u} \rho_S(x,t)$.

Equivalently, we may describe the shape by specifying, for all $n$, the
point $X_B(n,t)$ in price space above which exactly $n$ buyers may be
found, and the point $X_S(n,t)$ below which exactly $n$ sellers may be
found.  $X(n,t)$ is the inverse of the function $f(x,t) = \sum_{x'=0}^x
\rho(x',t)$.  The best bid $B(t)$ is then $B(t) = X_B(1,t)$, and the best
offer $O(t)$ is then $O(t) = X_S(1,t)$.  And we may interpret $X_B(n,t)$ as
the tender price necessary to make $n$ purchases (immediately), and the
price paid for the $n$ purchases would be $\sum_x^{X_S(n,t)} x 
\rho_S(x,t)$.

We may then express the density of traders near the best bid as $S_B$ as
$S_B(t) = \rho_B(B(t),t)$, and the density near the best offer $S_O$
similarly as $S_O(t) = \rho_S(O(t),t)$.  Of course, the bid-offer spread
is then ${\rm Spr}(t) = O(t) - B(t)$, and the midmarket is $M(t) =
(1/2)(B(t) + O(t))$.  Note that $B(t)$, $S(t)$, and $M(t)$ are not Markov
random variables, and do not satisfy a stochastic differential equation,
because they are subject to jumps.

Finally, the trading screen describes more than just the shape of the
reaction front, because it also describes the history of trades,
through the last trade price $\chi(t)$, and through the instantaneous
deal rate (trading volume/sec) $J$.  We show how these are
parametrized in the appendix.

{\bf Digression on the history of Diffusion-Controlled Annihilation
in the physical and chemical literature. }

The foregoing arguments have established the nature of our model,
commonly known in the physics and chemistry literature as steady state
diffusion- driven annihilation in one dimension.  We pause here to
mention previous work on this and related subjects.  As previously
mentioned, our model considers a steady state market, in which buyers and
sellers are injected into the system at a rate $J$ which precisely
compensates the losses due to annihilations .  However, one can also
consider the nonstationary situation in which no traders are injected to
compensate for the annihilations -- in this case the width of the
reaction front grows with time.  In the initial paper by G\'{a}lfi and
R\'{a}cz \cite{gr} the properties of the reaction front in a system with
segregated initial conditions were studied in the mean field
approximation.  The mean field approximation is one which approximates
the state by a Gaussian centered on the most heavily weighted
configuration, and ignores higher, non-Gaussian corrections, known as
``fluctuations''.  It works well for dynamics in three dimensional space,
but breaks down in lower dimensions due to the important role of
microscopic density fluctuations in one and two-dimensional systems and
in one dimension their influence is particularly important \footnote{ In
a series of papers Cardy et. al. \cite{cardy} shows how this obstacle may
be partially overcome, and we shall use his method in the Analytical
Results section}.  Numerical simulations were performed in
\cite{d1numerical}, and showed, among other things that reaction rates
are a Gaussian function of $x \in {\cal P}$.  Analytical calculations by
Cardy et. al. \cite{cardy} confirmed these numerical results.

Let us also  note the  paper by Bak, Paczuski and Shubik
 (BPS)\cite{BPS} where a  model based on  diffusion-controlled
 annihilation   was  first applied to the stock market.
  These authors introduce a series of models based on
diffusion-controlled annihilation as a route towards recovering the
observed Levy-Pareto "fat-tail" distributions which are said to
describe the medium term evolution of the stock-market.  The
diffusion-controlled annihilation process is modified by various forms
of sociological interaction, similar in philosophy to our bias model.
It is further modified by the presence of so-called "rational traders"
each with a different strategy, that set their bids and offers
according to their own individual expectations.  The rational traders
in this model are non-identical, whereas the non-rational ``noise''
traders are identical to one another, and so the pure noise trader
case, which they discuss at length, has the same dynamics as our
minimal model.  Buyers who are deleted from the market, however, are
re-inserted randomly as sellers, and vice versa.  The full BPS model,
from which they obtain their best results, gets its unique and
interesting dynamics from the various different (history dependent)
strategies which must wrestle with one another in order to find a
(possibly non-existent) equilibrium.  This model, while based on
diffusion-controlled annihilation, is quite different from our model,
and certainly more complicated.  Nevertheless, the success of a model
based on diffusion-controlled annihilation in reproducing observed
statistics of trade prices provides some encouragement for our rather
different, but related project of liquidity scaling laws.

\newsection{The Scaling Laws.}

We now proceed to a description of each of the research problems to
which we alluded in the introduction.  We write down an expression
which computes each scaling law in terms of the expectations of the
operators in the previous section.  Where necessary, we provide
background, or define additional terms and notation.  We write down
what qualitative intuition tells us about the limits of the scaling
law.  In addition, for a few of the simpler problems, dimensional
analysis is sufficient to compute the scaling laws directly, or a
scaling form (for the minimal model) and we include these formulas as
well.  In these cases, the consistency of intuition with the results
of dimensional analysis provides a satisfying and nontrivial check on
our method.  More generally, of course, these scaling laws will be
evaluated by numerical simulation and we will report on these results
in a future publication\cite{ftrnumerical}.

In the introduction it was mentioned that there are several other
problems beyond the liquidity scaling laws which we may investigate with
this model, leading to scaling laws beyond those discussed in the first
section.  We would like to draw the reader's attention to this point.
This is, in fact, one of the most appealing features of this arena of
research -- there is a very large and diverse set of financially
meaningful scalings that are easy to measure, easy to calculate, and
should only require a simple statistical model, such as the models
discussed in this paper, to make reasonably accurate predictions.  As
with the liquidity scaling laws, there is an obvious qualitative
intuition which we may use as a guide and sanity check.

Furthermore, these additional scalings are not empty mathematical
abstractions.  Their qualitative behavior is readily apparent to market
practitioners, and they are readily measured -- they are correlations of
quantities such as the best bid, best offer, trader density
\footnote{this quantity can not usually be read off of a trade screen
reliably, but may be observed by active intervention into the market.  },
trade price, etc.  which are directly observable in the marketplace.

{\bf Liquidity Scaling Laws}

Liquidity in markets is measured in several different ways, using several
different quantities, as proxies.  The most common are the deal rate $J$, 
the time to
midmarket sale, the bid-offer spread, and the trader densities near the
best bid/offer.  As these quantities all measure the same thing, they
must all be monotonic functions of one another.  We list these relations
here, expressed as expectations of field operators discussed in the
previous section, evaluated in the measure which is the steady state
solution of our dynamics.  This state measure depends on the deal rate
$J$, and we include whatever information might be inferred about this
dependence from dimensional analysis.

\begin{itemize}
\item Bid Offer Spread
\bq
{\rm Spr} = < O(t) - B(t) >
\eq
\item Trade price variance
\bq
w^2 = <\chi^2(t)> - <\chi(t)>^2
\eq
\item Fluctuations in the Bid-Offer Spread
\bq
(\Delta {\rm Spr})^2 = < (O(t) - B(t))^2 > - < O(t) - B(t) >^2
\eq

We expect both ${\rm Spr}(J)$ and $w^2 \longrightarrow 0$ as $J
\longrightarrow \infty$.  Both the bid offer spread and $w$ have
dimension of {\it dollars}, so dimensional analysis suggests ${\rm
Spr}(J) \sim \sqrt{D/J}$, $w(J) \sim \sqrt{D/J}$ consistent with
intuition.  Using the results of \cite{cardy} we shall show in the next
section that there is a logarithmic correction to the the results of
dimensional analysis, so that we have $w(J) \sim J^{-1/2} \ln J$ for
finite size price space.  Cardy's method will further enable us to
calculate the auto-correlation function of trade prices $W(\tau) =
<\chi(t + \tau)\chi(t)> - <\chi(t)><\chi(t+\tau)>$.  In addition, it is
possible, and possibly quite informative, to measure the fluctuations in
the spread $\Delta {\rm Spr}$.  This has the same dimensions as the
spread itself, and so the same scaling law.

\item Time to MidMarket Sale/Purchase (with certainty $1 - \epsilon$)

\bq
{\tau}_S = g(1-\epsilon); ~~ f(\tau) = <J^e_{\rm cum}(t+\tau)
\Psi^{\dagger}_e(M(t),t)>;~~  g(f(x)) = x
\eq

$\Psi^\dagger_e(M(t),t)$ inserts a non-diffusing trader at the mid
market, who trades without ever leaving the market, i.e. he annihilates
his counterparty, but not himself.  $J^{(e)}_{\rm cum}(t)$ is the
cumulative deal rate operator, which computes the number of deals
transacted with the trader $(e)$ during time interval $t$ (see Appendix).
This quantity evaluates the average time he must wait before he is hit or
lifted.  This is not the only way to compute $\tau_S$, however. Another
way to calculate $\tau_S$, for example, is to study the decay rate of an
excited state of a steady state market with a $\Psi^{\dagger}_e(M(t),t)$
inserted at time $t$:
\bq
d/dt \ln <\Psi^{\dagger}_e(M(t),t)> =
-(1/\tau_S)
\eq.

We expect $\tau_S \longrightarrow 0$ as $J \longrightarrow \infty$.
Because the dimension of $\tau_S$ is {\it sec}, dimensional analysis
suggests ${\tau}_S \sim J^{-1}$, consistent with intuition.

A similar quantity was considered in the framework of a continuous price
space by Cohen-Maier-Schwartz-Whitcomb \cite{CMSW81}, who showed that the
probability of a limit order trade does not go continuously to 1 as a
putative limit order bid approaches the best offer. This came about
because the random process for the best offer was taken to be a finite
sum of $N$ Poisson processes, approaching a Wiener process in the limit
$N \longrightarrow \infty$.  Under these assumptions, the anomalous
probability behavior, and the existence of a non-infinitesimal bid-offer
spread was proved, which however vanishes in the Wiener limit.  Because
our traders are considered to readily change their bid and offer prices,
the analysis of Cohen-Maier-Schwartz-Whitcomb would require $N$ very
large, near to the Wiener limit, leading us to expect the probability of
trading to go to $1$ as a limit order bid is placed near to the best
offer.  In both models, $\tau_S \longrightarrow 0$ as $J \longrightarrow
\infty$.

\item Density near the best bid/offer as a function of deal rate

For a small intervals or price space $\Delta p$, $\Xi_B(x,t)$ represents
the number bids we can hit in the vicinity of the price, between $B(t)
+ x$ and $B(t)  + x + \Delta p$
\bq
\hat \Xi_B(x,t) = \rho_B(B(t)+ x,t) \ \ \ \ \hat \Xi_O(x,t) = \rho_O(O(t) +
x,t)
\eq
\bq
\Xi_B(x,t) = <\hat \Xi_B(x,t)> \ \ \ \ \Xi_O(x,t) = <\hat \Xi_O(x,t)>
\eq
This quantity provides to us information about the ease with which we can
move the market, and the transaction costs of trading in large volume all
at once.  We expect $\Xi_B, \Xi_O \longrightarrow \infty$ as $J
\longrightarrow \infty$.  Dimensional analysis leads to the scaling law
$\Xi_{(\cdot)}(0) \sim (J/D)^{1/2}$, and $\Xi_{(\cdot)}(\Delta x)$ has
the scaling form $\Xi_{(\cdot)}(\Delta x) = (J/D)^{1/2}f(J \Delta
x^2/D)$.

\end{itemize}

The expressions above are the mathematical formulation of the qualitative
laws discussed in the introduction.  In a future publication, we shall go
beyond simple dimensional analysis, and report on results from actual
numerical simulations of the model\cite{ftrnumerical}.

{\bf Higher Correlation Functions Describing Equilibrium}

Beyond scaling laws directly associated with liquidity, there are many
others which explore the dynamical behavior of an equilibrium market.
These quantities are normally associated with correlations of operators
at different times, and/or conditioning of the expectation.  The time
differences $\Delta t_i$ between the different operators,and the
conditioning parameters now enter our calculations as new dimension-ful
parameters.  This weakens the power of dimensional analysis, because
there is no longer a single, unique dimension-ful combination of
parameters which matches the dimensions of the operator we are averaging.
Thus, for example, correlation functions will have different $J$
dependence for small $\tau << 1/J$ and large $\tau >> 1/J$.  One may
consider following functions:

\begin{itemize}
\item Correlation of Changes in Best Bid with Changes in Best Offer

For any quantity $f(t)$, define $\Delta_\tau f(t) = f(t+\tau) - f(t)$.
\begin{equation}
C_{\Delta_t B\; \Delta_t O}(\tau) =
{ <(B(t + \tau) - B(t))(O(t + \tau) - O(t))> \over \sqrt{
<(B(t + \tau) - B(t))^2><(O(t + \tau) - O(t))^2>} }
\label{C_bo}
\end{equation}
When the best bid snaps back after a trade, it moves down.  When, at the
same moment the best offer snaps back it moves up.  Afterwards, the best
bid resumes diffusing upward, while the best offer then diffuses back
downward.  We therefore expect that, if we look forward only over time
scales much shorter than one dealing time,$\tau_D \sim 1/J$,
\begin{equation}
C_{BO}(\tau) \longrightarrow   { < dB(t)/dt \; dO(t)/dt)>
\over \sqrt{ <dB(t)/dt>^2 \; <dO(t)/dt>^2 } } \longrightarrow -1
\end{equation}
Dimensional analysis suggests that $< dB(t)/dt\; dO(t)/dt)> \sim - D J$

But when we look forward over many dealing times, the market will
have moved to one centered around a completely new midmarket,
and the best bid and best offer will follow, together.  For these
time scales $\tau >> 1/J$,  $C_{BO}(\tau) \longrightarrow 1$, if that
new midmarket is centered sufficiently far away.

\item Density at/near Best Bid/Offer conditioned on
 Fluctuations in Best Bid/Offer:
\bq
\Xi_B(\sigma_B) =
{<{\rho}_B(B(t),t) I((B(t) - \bar B)^2  = \sigma^2_B)> \over
< I((B(t) - \bar B)^2  = \sigma^2_B)>} \nonumber
\eq
\bq
= E( {\rho}_B(B(t),t) | (B(t) - \bar B)^2  = \sigma^2_B )
\label{X_b}
\eq
$I(\cdot)$ is the indicator function.  (For simplicity, we discuss only
the bid -- the offer side case is, of course, identical).  If the density
of buyers is higher, then we can expect that the best bid does not need
to snap back as far when it jumps back to the next best bid.  We
therefore expect that $\Xi_B$ will be small at large $\sigma_B$, and
large at small $\sigma_B$. As $\sigma_B$ represents an additional
dimension-ful parameter, with dimensions of {\it (dollars)}, we cannot
use dimensional analysis to infer the $J$ dependence of
$\Xi_B(\sigma_B)$, however we do know that it will be of the form
$\Xi_B(\sigma_B) = \sqrt{J/D}f({D \over J \sigma^2_B})$.  Here $f$ is an
unknown function of a {\it single} variable, whose expansion in its
argument must contain only negative powers.  We may generalize this
quantity to explore the density beyond the best bid/offer, as follows.
\bq
\Xi_B(\sigma_B,\Delta x) =
{<{\rho}_B(B(t) + \Delta x,t) I((B(t) - \bar B)^2  = \sigma^2_B)> \over
< I((B(t) - \bar B)^2  = \sigma^2_B)>}
\label{X_b2}
\eq
This is a more complicated quantity, whose Laurent expansion coefficients
may be treated in a similar way.

\item Time Change in Spread conditioned on Spread:
We define the conditional expectation
\bq
\Delta_{\tau} {\rm Spr}(s_0) =
{<( \Delta_{\tau} {\rm Spr}(t) I({\rm Spr}(t) = s_0 )> \over
< I({\rm Spr}(t) = s_0 ) > }
= E( \Delta_\tau {\rm Spr}(t) | {\rm Spr}(t) = s_0 )
\eq
where $I(\cdot)$ is the indicator function.  The spread in a double
auction market starts out wide, then narrows slowly until the bid and
offer meet, at which point a deal is made and the best bid and offer snap
back.  Thus the expected change in time of the spread ($ \Delta {\rm
Spr(t)}{\rm Spr}(t + \tau) - {\rm Spr}(t)$) conditioned on the value of
the spread is small and negative for large values of the spread, and
large and positive for small values of the spread.  The zero point of
this function may be interpreted as the average width at which a deal 
becomes
imminent (the price is ``irresistible'').  When $\tau$ is close to zero,
so that the expectation is a derivative, there is a scaling form for this
quantity as well -- $\Delta_t {\rm Spr}(s_0) \sim \sqrt{D/J} f({D \over J
s_0^2 })$, where again $f$ is an unknown function of a single variable.
Because of our intuition that $\Delta_t {\rm Spr}$ increases sharply as
the value $s_0$ of ${\rm Spr}$ decreases, it is reasonable to expect that
the unknown function $f(x) \longrightarrow
\infty$ as $x \longrightarrow \infty$.

{\bf Market Response to a Disequilibrium of Buyers and Sellers}

When shocked into a state of disequilibrium, a market will go through
changes to its internal structure as it finds a new equilibrium.  We
may study the characteristic rates and magnitudes of these changes, to
each of the several parameters describing an equilibrium market.

\item Response Functions to Density Imbalance:
We define the conditional expectations
\bq
R_B(t,\tau) =
	{< \Delta_\tau B(t) I( \rho_O(O(t),t) - \rho_B(B(t),t) = \Delta 
\rho)>
\over < I( \rho_O(O(t),t) - \rho_B(B(t),t) = \Delta \rho)>} \nonumber
\eq
\bq
 = 	E (\Delta_\tau B(t) | \rho_O(O(t),t) - \rho_B(B(t),t) = \Delta
\rho) \nonumber
\eq
\bq
R_O(t,\tau) =
       { < \Delta_\tau O(t) I(\rho(O(t),t) - \rho(B(t),t) = \Delta
\rho)> \over
 < I(\rho(O(t),t) - \rho(B(t),t) = \Delta \rho)>} \nonumber
\eq
\bq
R_J(t,\tau) = {<\Delta J(t) I(\rho(O(t),t) - \rho(B(t),t) = \Delta
\rho)> \over
< I(\rho(O(t),t) - \rho(B(t),t) = \Delta \rho)>}
\eq
where $I(\cdot)$ is the indicator function.  From these response
functions we may derive characteristic parameters for the market's
response to a small disturbance, such as the response time $\tau_{R}$
to a density imbalance, or the expected size of the movement, of any
one of the response functions, e.g. $R_B$, due to a density imbalance
$\Delta R_B =  R_B(t,\tau) - \lim_{\tau \rightarrow 0} R_B(t,\tau)$.  For
$\tau_{R}$ we expect that, as $J \longrightarrow \infty, \tau
\longrightarrow 0$, and dimensional analysis suggests that $\tau_{R}
\sim J^{-1}$ \footnote{In fact, as it is reasonable to expect a
response to density imbalance proportional to the magnitude, we may
expect an exponential behavior in the response function, and may in that
case define $\tau_{R}$ as $\tau_{R}^{-1} = -d/d\tau \ln R(t,\tau)$.  } .
For $\Delta R_B$, we expect that as $J \longrightarrow \infty$, $\Delta
R_B \longrightarrow 0$, and as $\tau \longrightarrow \infty$, $\Delta R_B
\longrightarrow {\rm finite}$.  In addition, as $\tau \longrightarrow 0$,
this quantity becomes a derivative times $\tau$, so that $\Delta R_B \sim
\tau \sqrt{DJ}$, and as $J \longrightarrow 0$, $\Delta R_B \longrightarrow
\infty$.  Dimensional analysis suggests the scaling form $\Delta R_B
\sqrt{D/J} f(J\tau)$, with $\lim_{x \longrightarrow \infty} f(x) = {\rm
finite}$, and
$\lim_{x \longrightarrow 0} \sim x$.
$\Delta R_B \sim J^{-1}$.  Of particular interest is the final resting
place of the best bid, $\Delta R_B(\tau = \infty)$.

{\bf Transaction Costs }

\item Expected Profits for a Specialist:
Given a Bid Offer Spread set by a specialist, what is the expected
profit or loss?

This problem has been studied in the context of market makers by
\cite{Garman,Stoll,HoStoll,OharaOldfield,CMSW78,Hamilton76,CHMSW}.  Here
we study the specialist problem, using our diffusion-annihilation model.
A specialist sets a bid and an offer price in the market to provide
liquidity, and makes a profit from the spread, just as a market maker
does.  However, he doesn not monopolize trading -- other limit order
traders may trade with each other.  This problem therefore retains the
difficulties of a double auction market and requires treatment by
statistical field theory.  A full treatment also requires consideration
of the effects of inventory, asymmetric information, and other
traditional aspects of the market maker problem.

A specialist sets the
width and midmarket of his quoted prices, in order to maximize profits
and minimize risk.  Define the net profit/loss as the sum
of all of netted out positions $P_N(t)$ after time $t$, and define a
second quantity to be the net open position $P_O(t)$ after time $t$.
\bq
f_i =J_{\rm cum}^{(e_i)}(t')
{\Psi}^\dagger_{B(e_1)}( M(t) - \Delta B,t)
\Psi^\dagger_{S(e_2)}(M(t) + \Delta S,t)
\eq
\bq
P_N = \min_{i =1,2} f_i ~~~~~~~~~~~~~~~ P_O = (f_2 - f_1)
\eq
$\Psi^\dagger$ inserts an external non-diffusing trader, who trades
without ever leaving the market, ie.e he annihilates his counterparty,
but not himself.  $J^{(e_i)}_{\rm cum}$ measures the total volume of 
trades for
the trader $(e_i)$.

We define a good market-making strategy as one that results in a positive
expected profit, and zero expected net open position, i.e. $<P_O(t)> \,
\, = 0$, $<P_N(t)> \, \,> 0$ for all $t$.  In addition, we may define the
fluctuations in both of these quantities $<P^2_O(t)> - <P_O(t)>^2$, and
$<P^2_N(t)> - <P_N(t)>^2$.  They define the two sources of uncertainty
for a market making strategy, and we may use them to measure the
risk/reward ratio of a particular strategy for setting the market maker's
bid and offer.

\item Tender Offer:

How far above the best offer $\Delta x_N$ must we bid in order to attract
a given number of sellers $N$ within a specified time $t_{T}$, with
certainty $1-\epsilon$?

The operator expectation we must construct is very similar to the one
in the item above.
\bq
f(\Delta x) = <J_{\rm cum}^{(e)}(t + \Delta t)
{\Psi}^\dagger_{B(e)}( O(t) + \Delta x,t) + \sum_{O(t)}^{O(t)+\Delta
x}\delta S \rho_B(O(t)+s)>
\nonumber
\eq
\bq
{\Delta X}_N(\Delta t) = f^{-1}(\epsilon)
\eq
$\Psi^\dagger_{B(e)}$ inserts an external non-diffusing trader who trades
once, and disappears from the market.  $J^{(e)}_{\rm cum}$ measures the
total value of the trades done through this trader.
Recall that $\delta S$ is the lattice spacing.  For $\Delta t =0$, this
quantity simplifies greatly, to just
\bq
{\Delta X}_N(\Delta t =0) = < X_B(N,t) >
\eq
which scales as $\sqrt{D/J}$.

We expect that, as $J \longrightarrow \infty$, ${\Delta X}_N
\longrightarrow 0$, and conversely, as $J \longrightarrow 0$,
${\Delta X}_N \longrightarrow \infty$.  We also expect that, when $\Delta
t \longrightarrow \infty$, ${\Delta X}_N \longrightarrow {\rm finite}$,
and $\Delta t \longrightarrow 0$, ${\Delta X}_N
\longrightarrow<X_B(N,t)>$,  and that $\Delta x_N$ is monotonic.

Dimensional analysis predicts that ${\Delta X}_N(\Delta t)$ has the
scaling form ${\Delta X}_N \sim \sqrt{D/J}g(J\Delta t)$, and this implies
that $g(x)$ is finite and monotonic in both the limits $x \longrightarrow
0$, and $x \longrightarrow \infty$.  Note that the consistency of these
limits is a nontrivial check on the dimensional analysis result.

\end{itemize}

{\bf  Analytical Results}

Thus far we have discussed the predictions of our model in terms of
computer simulations and dimensional analysis.  In addition, however,
many analytical approximation schemes also have been developed for
statistical field theory.  Although numerical results will be more
precise, these methods supplement numerical simulations of the model
by providing intuition.

Cardy et. al. \cite{cardy} has developed one such approximation using
 a method known as mean field theory, in the context of a model of chemical
reactions in one dimension.  Mean field theory replaces the
evolution equations for the operators with a partial differential
equation for the density configuration with the greatest probability
mass, ignoring others that may be contained within the state (known as
the ``fluctuations'' away from the mean field).  This  scheme is
particularly interesting because he retains some of he effects of the
fluctuations by deriving an effective noise term.  Mean field theory,
however, is not a systematic approximation scheme -- it has no system
of higher order corrections, and there is no known error estimate.
Furthermore, the  scheme allows us only to calculate a small subset
of the scaling laws of the previous section.  However, its predictions
coincide nicely with numerical simulations\cite{cardy,d1numerical}.

A model of chemical reactions must differ from a model of market
microstructure in certain respects.  First, the configuration space
for a chemical species can only be continuous, in contrast to our
discrete price space ${\cal P}$.  Also, chemical reactions differ from
deals in a marketplace, in that they occur with only a finite
probability per unit time $\lambda$ when the reactants have been joined,
whereas exchange rules guarantee an immediate trade when a bid
coincides with an offer.  Thus the chemical model has the following
time evolution operator
\bq
U_\lambda( t,t')  = \exp{ H_\lambda(t'-t) } ~~~~~~~~~~~~~~~~~~~~~~
\nonumber  \\
H_\lambda  = \int_{-L/2}^{L/2} dx
\left[\bar \Psi_b\left( - D\nabla^2 \Psi_b\right)
    + \bar \Psi_s\left( - D\nabla^2 \Psi_s\right)
  + \right.  \label{evolution} \\
 \left.  \lambda \left(\bar \Psi_b \bar \Psi_s -1\right) \Psi_b \Psi_s
  -\bar \Psi_b J \delta(x+L) -\bar \Psi_s J \delta(x-L) \right]
\nonumber
\eq
which may be derived from a master equation describing the evolution
of probability distributions (see for example \cite{cardy} and
references therein, see also the Appendix).  The first two terms describe 
diffusion of $B$
and $S$ particles, the next two terms describe annihilation $B + S
\rightarrow 0$ and the last two terms are necessary to insert $B$ and
$S$ particles at the edges of the system at the rate $J$.  The size of
the system is $L$, and our model corresponds to the limit $\lambda
\longrightarrow \infty$.  \footnote{ This operator differs from
that given in the appendix in that it corresponds to
\bq
\exp{\bigl(
H^{(B)}_{\rm Diff}(t-t') + H^{(S)}_{\rm Diff}(t-t') + H_{\rm
ann}(t-t') \bigr)},
\eq
instead of
\bq
\exp{\bigl( H^{(B)}_{\rm Diff}(t-t')
+ H^{(S)}_{\rm Diff}(t-t')\bigr)}\exp{ H_{\rm ann}(t-t')}.
\eq
Because the arguments of the exponential are operators, these are not the
same, and in particular, it not yet known whether this allows for some
overlap of buyers and sellers, even in the limit $\lambda
\longrightarrow \infty$, whereas it is explicit in the form given in
the appendix. }

In this section we shall outline Cardy et. al. method, and give several
results for the minimal model.  In later sections, where possible, we
shall also present results for some of the extended models.

In general, mean field theory leads to a system of scalar partial
differential equations.  These are obtained as the extrema of a
functional $S_\lambda$ of the fields $\bar \Psi_b(x,t)$, $\bar 
\Psi_s(x,t)$,
$\Psi_b(x,t)$ and $\Psi_s(x,t)$, $S_\lambda[\bar \Psi_b, \bar \Psi_s,
\Psi_b, \Psi_s]$, which is associated with the evolution operator
(\ref{evolution})
\bq
S_\lambda  = \int_{t'}^{t} dt\int_{-L/2}^{L/2} dx
\left[\bar \Psi_b\left({ \partial \Psi_b \over \partial t } -
 D\nabla^2 \Psi_b\right)
    + \bar \Psi_s\left({ \partial \Psi_s \over \partial t } -
D\nabla^2 \Psi_s\right)
  + \right.  \label{Lagrangian} \\
 \left.  \lambda \left(\bar \Psi_b \bar \Psi_s -1\right) \Psi_b \Psi_s
  -\bar \Psi_b J \delta(x+L) -\bar \Psi_s J \delta(x-L) \right]
\nonumber
\eq
These partial differential evolution equations can be obtained by varying
with respect to $\bar \Psi_{b,s}$ after a change of variables, the
redefinition $\bar \Psi_{b,s} \rightarrow \bar \Psi_{b,s} +1 $.
Neglecting the nonlinear term $\lambda \bar \Psi_{b}\bar
\Psi_{s}\Psi_{b}\Psi_{s}$ one gets the standard reaction rate equations
for the densities $\rho_{B,S}(x,t) = <\Psi_{b,s}(x,t)>$
\bq
(\partial_{t} - D\partial^2_{x})\rho_{B} +
\lambda \rho_{B}\rho_{S}- J\delta(x+L) =0; \nonumber \\
(\partial_{t} - D\partial^2_{x})\rho_{S} +
\lambda \rho_{B}\rho_{S}- J\delta(x-L) = 0
\eq
and one finds a particularly simple equation for the quantity
$\zeta(x,t) = \rho_{B}(x,t)- \rho_{S}(x,t)$, which may be interpreted
as the density difference of buyers and sellers.
$$
{ d\zeta \over dt } - D\nabla^2 \zeta = 0
$$
This equation is independent of the reaction rate $\lambda$, so the
limit $\lambda \longrightarrow \infty$ is obviated.  It can be shown
\cite{cardy} that to take into account fluctuations one has to take into
account the nonlinear term $\bar \Psi_{b}\bar \Psi_{s}\Psi_{b}\Psi_{s}$
which effectively adds a noise term in the right hand part of diffusion
equation
\bq
{ d\zeta \over dt } - D\nabla^2 \zeta = \eta
\eq
The equation retains some of the stochastic character of the original
system through the noisy source term $\eta$.  $\eta$ is a noise term,
satisfying $<\eta> = 0$, and
\bq
< \eta(x,t)\eta(x',t')> = 2R \delta(t-t')\delta(x-x')
\eq
Here $R$ is a function describing the local reaction rate at $x$, the
number of reactions (deals) per unit length per unit time, which has been
shown \cite{d1numerical} to have the scaling form $R(x) = (J/D) S(x/w)$,
where $w$ is the width of the reaction front.  After integrating $R$
over $x$ one has $\int dx R(x) = J$.  We seek solutions to this equation
with constant flux on the boundaries, whose noise-averaged value is zero
at the center of the front.  We enforce the condition that the
noise-averaged slope at the boundaries is $\sim -J/D$, so that the inward
flux of buyers and sellers at the boundaries is maintained at $J$ .  We
solve this equation on the closed interval $[-L/2, L/2] \subset {\cal
R}$, which here plays the role of the price space ${\cal P}$.  We write
the most general solution as an expansion in Fourier modes
\bq
\zeta(x,t) = -(J/D)x + \sum_{n=0} \chi_n(t) \cos\left({ (2n+1) \pi x
\over L} \right)
\label{zeta}
\eq
where the $\chi_n$ are the time dependent, noise dependent Fourier
coefficients.
\bq
\chi_n(t) = \frac{2}{L}\int_{0}^{t}dt'\int dx' \eta(x',t') \cos\left({ 
(2n+1) \pi x'
\over L} \right) \exp\left[\frac{ (2n+1)^2 \pi^2 D}{L^2} (t'-t)
\right]
\label{chi}
\eq
Modes $n$ with wavelengths ${L \over (2n+1)} < w$ oscillate many times
within the reaction front, and therefore do not contribute to the sum
(\ref{zeta}).  We may therefore simplify the calculation by approximating
the noise source $\eta$, distributed by the sharply peaked envelope
function $R$, as a point source at the origin,
\bq
\eta(t) = \frac{1}{\sqrt{2J}}\int dx' \eta(x',t') dx';
\eq
\bq
<\eta(t)\eta(t')> = \delta(t-t')\frac{1}{2J}\int dx R(x) =
\delta(t-t')
\label{eta}
\eq
and restricting the sum over $n$ to $0 < n < N_{\rm max} = {L \over w}$
\footnote{ The dependence on $w$ of our approximate expectation will lead
to a self-consistent definition for $w$, and a logarithmic correction the
dimensional analysis of the previous section}.

The quantity $\zeta(x,t)$ represents the difference between the
density of buyers and sellers at the point $x$, at time $t$.  This is
not directly one of the quantities we are interested in, but it is
related. The zero point of $\zeta$ represents the center of the
market, the midmarket $M(t)$, $\zeta(M(t),t) = 0$.  Since the density
difference goes through its zero point linearly with slope $\sim -J/D$,
its fluctuations at the point $x=0$ are proportional to the
fluctuations of the position of the zero of $\zeta$, i.e. the
midmarket.  Thus, the midmarket position $M(t)$ may be written
(neglecting terms of order $<M^2>/L^2 <<1$)
\bq
M(t) - \bar{M}  =
\frac{D}{J}~\zeta(0,t) = \frac{D}{J}~\sum_{n=0}^{N_{max}}~ \chi_n(t)
\label{midmarket}
\eq
where we have allowed for the possibility $\bar{M} = <M> \neq 0$ even
though $<M>=0$ in (\ref{zeta}).  The noise averaged value of $M(t) - \bar
M$ is $0$, but the noise dependent terms cause it to fluctuate, and we
may calculate the magnitude of these fluctuations
\bq
w^2 =  \frac{D^2}{J^2}(<\zeta^2(0,t)> -
<\zeta(0,t)>^2)  =
  \frac{D^2}{J^2} \sum_{n}^{N_{max}} \sum_{m}^{N_{max}}<\chi_n(t)\chi_m(t)>
\eq
as well as the autocorrelation function $W(T) = <M(t)M(t +T)> - <M(t)><M(t 
+T)> $
\begin{eqnarray}
W(T) =
 \frac{D^2}{J^2} ( <\zeta(0,t)\zeta(0,t + T)>- <\zeta(0,t)><\zeta(0,t +
T)>)  \nonumber
\eq
\bq
 = \frac{D^2}{J^2} \sum_{n}^{N_{max}}
\sum_{m}^{N_{max}}<\chi_n(t+T)\chi_m(t)>
 \label{correlatorW}
\end{eqnarray}
The angle brackets  mean average over the noise $\eta$,
invoking the expression (\ref{eta}) above. Using (\ref{zeta}) -
(\ref{eta}) it is easy to see that for large $t$
\bq
<\chi_{n}(t)\chi_{m}(t)> =
\frac{8J}{\pi^2D}\frac{1}{(2n+1)^2 + (2m+1)^2}
\label{chichi}
\eq
and one gets \cite{cardy}
\bq
w  =  \left[ {\ln(cL/w) \over \pi (J/D) } \right]^{1/2}
\label{width}
\eq
where the logarithmic factor is due to the summation $\sum_{n,m}[(2n+1)^2
+ (2m+1)^2]^{-1}$ over $n$ and $m$ from $1$ to $N_{max} = L/w$.  As
expected from dimensional analysis, the width of the reaction front goes
like $\sqrt{D/J}$.  Note also that the midmarket variance grows, slowly,
with $L$.  This is an example of the logarithmic corrections to our crude
dimensional analysis of the scaling laws mentioned in section 2.  It is
interesting because it implies that having broad masses of traders behind
the best bid and offer will have a (weak) influence on the behavior of
the midmarket, increasing its variance -- presumably other scaling laws
will be similarly affected.  Thus, choosing the parameter $L$ in our
model is not an arbitrary choice, or technical problem.  $L$ defines how
tightly traders are bunched together in price space outside of the
reaction front, and is therefore a parameter carrying meaningful
dynamical information about the market, which we may measure along with
the other parameters of the market, $J$ and $D$.

Finally we note that the width (\ref{width}) does not widen with time,
which makes sense in a stable market, not subjected to external shocks.
We shall see in the next section how these shocks actually may be added
to our model, resulting in the usual result for Brownian motions of a
variance growing linearly in time.

We may calculate the correlator (\ref{correlatorW}) by the same method as
the width.  The result is
\begin{eqnarray}
W(T) = {2D\over J \pi^2} \int {dx dy \over x^2 + y^2}
e^{-x^2 J |T|} \approx
{D\over J \pi}\left( E_1(J|T|) - E_1(w^2 J |T|/L^2)\right)
\end{eqnarray}

where the integration is from $\epsilon = \sqrt{D/JL^2} <<1$ to 1 and
$E_1(x)$ is the exponential integral $E_1(x) = \int_x^\infty { e^{-t}
dt \over t}$.  For small $x$ $E_1(x) \sim -\ln x$ and for large $x$ it
is exponentially small.  There are three time intervals: $T < 1/J, ~
1/J < T < 1/J\epsilon^2$ and $ T> 1/J\epsilon^2$.  In the first
interval $x^2 J |T| < 1$ for all $x$ and we can expand $e^{-x^2 J |T|}
\approx 1 - x^2 J |T|$ which gives us $W(T) = w^2 - (1/2\pi) DT$.  In
the interval $1/J < T < 1/\epsilon^2 J$ one can neglect $E_1(J|T|)
\sim e(-JT)$ and only $ E_1(\epsilon^2 J |T|)$ is important, so
\begin{eqnarray}
W(T) \approx {D\over 2\pi J} \ln\left[\frac{L^2}{w^2 J T}\right],~~~
1< J T < 1/\epsilon^2
\end{eqnarray}
In the third interval the correlation function is exponentially small
$W(T) \sim e(-(w^2 J |T|/L^2))$ - but $T > 
1/\epsilon^2 J$ represents
time scales so long as to be beyond the scope of this model.

Note that, because the correlation function is a function of the time
difference $T$ a measurement of $W(T)$ for a range of values $T$ contains
enough information to fix all of the unknown parameters in the minimal
model.

{\it Nonstationary Markets: the Minimal Model with Asymmetric Fluxes }

It is also possible to consider the minimal model with different fluxes
for the buyers and the sellers.  Cardy et.al. scheme may be used with few
modifications. The principal difference is the non-stochastic part of the
solution $\zeta(x,t)$ must be time dependent, and satisfy the appropriate
slope conditions at the boundaries.  Let $\bar J \pm \Delta J$ be the
flux at the upper (lower) boundary.  Then the most general solution for
$\zeta$ is now
\bq
\zeta(x,t) = -{\bar J \over D}x -{\Delta J \over 2LD } x^2 - {\Delta J 
\over
L }t + \sum_{n} \chi_n(t) \cos\left({ (2n+1) \pi x \over L} \right)
\eq
The nonstochastic part no longer vanishes at $x=0$ -- instead its zero is
a function of time $x_0(t)$.  For small times
\bq
x_0(t) \sim {-2D \Delta J \over \bar J L }t,
\label{MktVel}
\eq
and this defines the speed of the moving midmarket.
Following the same logic as before  we must then evaluate
\bq
w^2 =  \frac{D^2}{\bar J^2}\left( <\zeta^2(x_0(t),t)> - 
<\zeta(x_0(t),t)>^2\right) .
\eq
which leads to the result
\begin{eqnarray}
w_{\Delta J}^2(t) = {2D\over \pi^2\bar J } \int {dx dy \over x^2 + y^2}
 \cos( \nu x ) \cos( \nu y )
\label{wdeltaJ}
\end{eqnarray}
where $\nu = (2/\pi)(\Delta J/ \bar J) (\bar J t) \epsilon$
 Now $w_{\Delta J}$ explicitly depends on $t$
and because $ \cos( \nu x ) <1$ the
width is smaller and  decreasing with time.
For short times, this may be expanded out as
\bq
w_{\Delta J}^2(t) = w^2 -  {2D\over \pi^2\bar J }\nu^2
\eq
The correlation function may be calculated as well. The
result is
\begin{eqnarray}
W_{\Delta J}(t, t') = {\bar J \over D} \int {dx dy \over x^2 + y^2}
e^{-x^2 \bar J |t-t'|} \cos( \nu x ) \cos( \nu' y )
\end{eqnarray}
and now it is not only a function of $T = t-t'$, but also of $t$ and
$t'$ separately due to $ \cos( \nu x )$ and $\cos( \nu' y )$ factors.
One can again study the behaviour of the correlation function in detail
as before, and we do so in the next section.

\newsection{ Beyond the Minimal Model }

{\bf Markets Under the Influence of Random Disturbances}

As we discussed in the introduction, it is possible to consider the
minimal model with the additional feature of random disturbances applied.
This will have the effect of injecting ``energy'' into the market,
causing the system to be partly in the low lying excited states. Our
approach to calculating this sort of effect will be to introduce an
external source of noise into one of the parameters of the model, and
average our calculations over that noise source. Implicit within this
procedure is the assumption that these fluctuations occur over time
scales much shorter than those of measurement. We may
imagine several kinds of disturbances.

{\it Random Fluctuations in Trader Drift}

The first might be considered a disturbance due to the influence of news.
A market's internal structure is buffeted throughout the day by news
announcements, and the behavior of markets for related assets.  To model
this effect and observe its effect on the scaling of market
microstructure, we subject our diffusing traders to a random white-noise
drift $v(t)$, proportional in strength to the trade price
$<v(t) v(t')> = ST_d\delta(t-t')$, where $S$ is the approximate
price at the midmarket, and $T_d$ is a parameter which describes
the strength of the noise.  This noise is intended to be the response of
traders to a series of small news items, both good and bad.  The effect
of this on the stable market represented by the minimal model is to cause
the trade prices to execute a random walk.  We then modify
the equation for $\zeta$ as
\bq
{ d\zeta \over dt } - D\nabla^2 \zeta = \eta + v(t) { \partial \zeta \over
\partial x }
\eq
We can solve this equation by perturbation in $v(t)$, so
that $\zeta(x,t) = \zeta_{0}(x,t) +\zeta_{1}(x,t) +... $,
$\zeta_{0}(x,t)$ is a solution to (\ref{zeta})  and
\bq
{ d\zeta_{n} \over dt } - D\nabla^2 \zeta_{n} = \eta + v(t) { \partial 
\zeta_{n-1} \over
\partial x }
\eq
Because we are interested in finding $\zeta(x,t)$   at small $x$ it is
easy to see  from
\bq
 {\partial \zeta_{0} \over \partial x } = -(J/D) -
 \sum_{n=0} {(2n+1) \pi  \over L} \chi_n(t)
\sin\left({ (2n+1) \pi x \over L} \right)
\eq
 that one can neglect all $ \chi_n(t)$ terms, because they are
suppressed as $x/L$, so $ \zeta_{1}(x,t) = -(J/D) \int dt v(t)$ and
$\partial_{x} \zeta_{1} =0$, which means that
 the solution in the presence of drift is
\bq
\zeta(x,t) \approx -(J/D)x + \sum_{n=0} \chi_n(t) \cos\left({ (2n+1) \pi x
\over L} \right) - (J/D) \int dt v(t).
\eq
We see that the effect on the solution $\zeta$ is only to add a Brownian
motion to the solution, independent of $x$.
  This leads to the following results
for the reaction front width $w^2 = <M^2> - <M>^2$ and correlation function
 $W(t-t') = <M(t)M(t')> - <M(t)><M(t')>$  for  the MidMarket
field $M(t) = (D/J) \zeta(0,t)$

\bq
 w^2(t) = { D \over\pi  J} \ln (1/\epsilon)  + S T_d (t-t_0) \nonumber  \\
W(t, t') =  W_{0}(t-t') + S T_d \min((t-t_0),(t'-t_0))
\eq
and $t_0$ is the moment when the drift $v(t)$ was switched on.  The
fluctuations in $M(t)$ start out as a constant, but then after the
characteristic time ${D \over JST_d}$, they grow in time, as expected
for a Brownian motion.  This is consistent, both with standard option
theory, and with the common observation that intra-day volatilities
are larger than expected for a Brownian motion.  Furthermore, we see
that the parameter $T_d$ should be identified with the volatility of
the trade price.

{\it Random Fluctuations in the Flux of New Traders}

The second type of noise to which we subject the system is a fluctuating
rate of traders entering the market.  It is obviously somewhat more
general to assume a randomly fluctuating entry rate than a constant one,
and so, at the expense of introducing an additional parameter, we nudge
our model one step closer to a realistic market.  Following Cardy, we may
once again compute mean field theoretic approximations to some of the
market microstructure.  This calculation may be done directly from the
results of the previous section, for a market with unequal trader
influxes, by taking the quantity $\Delta J$ to be randomly fluctuating
$<\Delta J(t)> =0, <\Delta J(t) \Delta J(t')> = \mu^2 \bar J^2
F(t-t'), ~F(0) =1$. Note that we must have $\mu << 1$ in order to ensure
that $J$ is always positive.  Using (\ref{wdeltaJ}) we find
\begin{eqnarray}
w_{\Delta J}^2(t) = <{2D\over \pi^2\bar J } \int {dx dy \over x^2 + y^2}
 \cos( \nu x ) \cos( \nu y )>
\end{eqnarray}
We calculate the average using the relation
\bq
<\cos( \nu x ) \cos( \nu y
)> = 1/2[\exp(- <\nu^2> (x+y)^2/2 ) + \exp(- <\nu^2> (x-y)^2/2 )]
\eq
and making use of $\nu = (2/\pi)(\Delta J/ \bar J) (\bar J t) \epsilon$
we have $<\nu^2> = (2\bar J \epsilon /\pi)^2 \mu^2 t^2$.  Expanding in
$<\nu^2>$ we have for small $t$
\begin{eqnarray}
w_{\Delta J}^2(t) \approx {2D\over \pi^2\bar J }
 \int {dx dy \over x^2 + y^2}
 \left[1 - \frac{1}{2} <\nu^2>(x^2+y^2)\right] \nonumber \\
= \frac{D}{\pi \bar J}\left[ \ln(1/\epsilon) - \frac{ (2\epsilon\mu
\bar{J} t )^2}{\pi^3}\right]
\label{diffsncrxn}
\end{eqnarray}
and we see by solving (\ref{diffsncrxn}) that the correction to the total
fluctuation width due to random noise in $J$ is negative, and quadratic in
time.

{\it Random Fluctuations in Trader Volatility}

Finally, we consider noise added to the trader volatility $D$, and
again we may calculate its effects in mean field theory.  To this end,
we introduce a small noisy term $d(t)$ to be added onto the average
diffusion constant $\bar D$, $D = \bar D + d(t)$, where $<d(t)> = 0$
and $<d(t)d(t')> = T_D\delta(t-t')$, and solve the equation
\bq
{ d\zeta
 \over dt } - D\nabla^2 \zeta = \eta + d(t)\nabla^2 \zeta
\eq
perturbatively in $d(t)$: $\zeta = \zeta_{0} +\zeta_{1} +...$.  $T_D$
must be very small $T_D << \bar D^2$, so that the total trader volatility
$D$ is always positive.  The equation for $\zeta_{1}$ is
\bq
{ d\zeta_1 \over dt } - \bar{D}\nabla^2 \zeta_1 =  d(t)\nabla^2 \zeta_0
  \nonumber \nonumber
\eq
\bq
= -  d(t) \sum_{n}\chi_{n}(t)\left({ (2n+1) \pi \over L} \right)^2
\cos\left({ (2n+1) \pi x\over L} \right)
\eq
and
\bq
\zeta_1(x,t) =  \sum_{n=0} \chi^{1}_n(t) \cos\left({ (2n+1) \pi x
\over L} \right)
\label{zeta1}
\eq
where $\chi^1_n$ is easily found to be
\bq
\chi^1_n(t) = - \int_{0}^{t}dt' d(t')\chi_{n}(t')
\left({ (2n+1) \pi \over L} \right)^2
 \exp\left[\frac{ (2n+1)^2 \pi^2 \bar{D}}{L^2} (t'-t)
\right]
\label{chi1}
\eq
and $\chi_{n}(t')$ is given by (\ref{chi}). It is easy to see that
$<\zeta_{0}(t) \zeta_{1}(t)> = 0$ because $\zeta_{0}(t)$ does not
depend on the noise $d(t)$ and $\zeta_{1}(t)$ is linear in $d(t)$ so
$<\zeta_{0}(t) \zeta_{1}(t)> \sim <d(t)> = 0$.  Then $<\zeta^2(t)> =
<\zeta_{0}^2(0,t)> + <\zeta_{1}^2(0,t)>$ and the midmarket fluctuation
width is given by
\bq
w^2 = <M^2(t)> = \frac{D^2}{J^2} \sum_{n,m}\left[<\chi_{n}(t)\chi_{m}(t)>
 + <\chi^1_{n}(t)\chi^1_{m}(t)> \right]
\eq
We have already calculated $<\chi_{n}(t)\chi_{m}(t)>$ (\ref{chichi})
and using the same method it is easy to calculate
\bq
<\chi^1_{n}(t)\chi^1_{m}(t)> = T_D\frac{\pi^2}{L^2 \bar{D}}
\frac{(2n+1)^2(2m+1)^2}{(2n+1)^2+(2m+1)^2}<\chi_{n}(t)\chi_{m}(t)>
\nonumber  \\
= T_D\frac{8J}{L^2 \bar{D}^2}
\frac{(2n+1)^2(2m+1)^2}{[(2n+1)^2+(2m+1)^2]^2}~~~~~~~~~~~~
\eq
After summation over $n$ and $m$  we get an extra factor
$N_{max}^2 = L^2/w^2$ and the width
\bq
w^2 = {\bar{D} \over \pi J}\ln(1/\epsilon)  + {\pi \over 4} {T_D \over
J}{1 \over  w^2}
\eq
and we see that the correction scales as $1/J w^2$ for small $T_D$, and
is independent of time.  When $T_D$ gets larger the scaling law for $w^2$
is going to change - it is clear that for $J > \bar{D}^2/T_D$ the second
term is bigger than the first one and one might conclude that $w \sim
J^{-1/4}$.  However, $T_D$ cannot be too large, because otherwise the
total trader volatility $D = \bar D + d(t)$ can go negative.

In conclusion let us note that the three different fluctuations in the
market parameters lead to three different types of corrections to the
width: fluctuations in $J$, $D$, and the trader drift parameter each lead
to corrections to the width.  Noise in the drift leads to a positive
correction to $w^2$, growing linearly in $t$, and independent of $J$.
Noise in $J$ leads to a negative correction, growing quadratically in
time, and noise in $D$ leads to a positive correction independent of
time. The corrections to the autocorrelation function are also
calculable, and are interesting because the function contains more
information, which we may use to extract the parameter of $D$, $J$, and
$L$ precisely, and to provide additional tests of these conjectures.

{\bf Two-Liquid Model}

During a crash, it is clear that a market behaves altogether differently
from the quiescent markets on which our efforts have so far been
concentrated.  It is known, for example, that bid-offer spreads widen out
hugely.  Nevertheless, it is reasonable to expect a market in this state
to exhibit {\it some} sort of scaling behavior.  In this picture, the
crashing market is a different phase, as a gas phase is different from a
liquid phase.  A model for this phase would have to have a different
ground state, one which allowed for this sort of crashing dynamics.

The simplest model for a market crash was suggested at the end of the
minimal model discussion, in which a mean field calculation was done for
some of the market microstructure when buyers and sellers entered the
market at unequal rates.  However, this does not cause spreads to widen
out appreciably, and there is a good reason why.  Best bids and offers
can only be calculated from the bids and offers of limit-order traders,
who submit specific prices at which they will trade.  During a crash, a
large number of trades are executed as market orders, that is, with no
prior warning of the position in price space that that trader had
occupied.  We thus introduce a second type of buyer and seller into the
market, the market order trader, who is distinguished from the limit
order trader only in that we cannot see his bids and offers on screen.
Thus, the best bid and offer continue to be calculated from the limit
order traders, even though most of the trading volume is now being
conducted by market order sellers, with limit order buyers.

This model can be formulated as one in which there are two flavors of
buyer, $B$ and $B'$, and two flavors of seller $S$ and $S'$.  $B$ and
$S$ are limit order traders and $B'$ and $S'$ are market order
traders.  The prices of the unprimed traders {\it only} are used to
calculate the best bid and offer on the trade screen, and are
therefore visible to other traders.  Thus the two-fluid model includes
the reactions $B + S \longrightarrow 0$, $B' + S \longrightarrow 0$,
$B + S' \longrightarrow 0$, but not $B' + S' \longrightarrow 0$,
because these primed traders are invisible to each other.  It is easy
to write an evolution operator like (\ref{evolution}) for all 4 fields
$\Psi_{b}, \Psi_{b'} , \Psi_{s}$ and $\Psi_{s'}$ which describes the
diffusion of all 4 fields and annihilation interactions between them
of the form $ \lambda \left(\bar \Psi_i \bar \Psi_j -1\right) \Psi_i
\Psi_j$ for three possible pairs $(i,j) = (b,s),~ (b',s)$ and $(b,
s')$.  There is no pair $(b',s')$ - market order traders do not see
each other.

For low density and flux of the market order traders, the dynamics of this 
model
is expected to be dominated by limit order traders, and is essentially
the same as the minimal model.  Consider now the case where the
density and flux for market order traders equals that of the limit
order traders.  In this case there are very few dimension-ful
parameters, so dimensional analysis regains some predictive power.  We
expect scaling laws exactly like those of the minimal model, in
particular the spread should scale like ${\rm Spr} \sim
\sqrt{(D/J)}$.  However, market order traders cause the spread to
widen out, because they trade only with limit order traders, pushing
the spread farther out each time they do.  This effect was discussed
in \cite{CMSW81}.  Thus, we expect a numerical prefactor, greater than
$1$, relating the minimal model spread to that of the two-liquid model
at this symmetric point.

When the density and flux of market order traders dominate that of the
limit order traders, we expect a very wide spread, as a result of the
market order traders pushing limit order bids and offers far away from
each other.  In this situation the market order best bid $B'$ is very
near to the limit order best offer $O$: $B_t' \sim O_t$, and is far
above the market order best offer, which is near to the limit order
best bid: $O_t' \sim B_t$, in the fashion of an across market, because
they cannot see one another.  In this case, the limit order traders no
longer trade with one another, because they are so far away, and we
have two decoupled minimal model systems, the $(b',s)$ system and the
$(b,s')$ system.  The averaged spread in this case grows linearly with
time, with each side moving at the rate determined by the formula
(\ref{MktVel}) from the previous section.  In practice, as discussed by
\cite{CMSW81}, this unstable situation stabilizes itself when market
order traders switch themselves to limit order traders, but that is
beyond the scope of this model.  Finally, in a crash situation, we
expect an asymmetric flux of buyers and sellers, with most sellers
being market order sellers.  Again the $(b',s)$ and the $(b,s')$
systems are decoupled into two minimal models, and the averaged spread
again grows linearly with time, as in formula (\ref{MktVel}).  In this
situation, which is the situation of interest in this model, market
order sellers cannot switch to limit order sellers and reasonably
expect to get their order executed, so this result is expected to
hold.

This model is much harder to treat analytically -- method of Cardy
 and collaborators \cite{cardy} may
be used, but none of the quantities accessible to us are interesting.
We have 4 fields instead of 2 and many more free parameters (different
fluxes for limit and market order traders, etc.).  In the mean field
approximation, neglecting the quartic term, one gets for the densities
$\rho_{i}(x,t) = <\Psi_{i}(x,t)>$
\bq
(\partial_{t} - D\partial^2_{x})\rho_{B} +
\lambda \rho_{B}(\rho_{S} +\rho_{S'})   =0;~~
(\partial_{t} - D'\partial^2_{x})\rho_{B'} +
\lambda \rho_{B'}\rho_{S} = 0;
 \nonumber \\
(\partial_{t} - D\partial^2_{x})\rho_{S} +
\lambda \rho_{S}(\rho_{B} +  \rho_{B'}) = 0;~~
(\partial_{t} - D'\partial^2_{x})\rho_{S'} +
\lambda \rho_{S'}(\rho_{B} = 0
\eq
and if all diffusion coefficients are the same $D = D'$ and one finds
a particularly simple equation for the quantity $\zeta(x,t) =
\rho_{B}(x,t) + \rho_{B'}(x,t) - \rho_{S}(x,t) - \rho_{S'}(x,t)$,
which may be interpreted as the density difference of all buyers and
sellers.  $$ { d\zeta \over dt } - D\nabla^2 \zeta = 0 $$ This is the
same mean field equation as in the minimal model but unfortunately the
non-linear effects cannot be incorporated through a single noise
term.  Furthermore, the zero of $\zeta$ cannot be interpreted as the
midmarket.  We are thus confined to numerical simulations for this
model, at least for now.  We shall report on computer simulations of
this ``Two Fluid Model'' in the crashing phase, in an upcoming
publication. It will be interesting to see if in the crashing phase
this model  leads to   log-periodic oscillations \cite{log}.  

{\bf The Bias Model}

As mentioned in the introduction, the bias model is a modification of the
minimal model introduced above, in which the diffusion operators $D_B$
and $D_S$ are modified to contain a drift, which increases when the last
trade has ticked upwards, and decreases when the last trade has ticked
downward.  (The amount by which it increases or decreases is controlled by
a parameter which we refer to as the ``market tension''.)  This means that
our state, our probability distribution at time $t$, now depends not only
on the positions of traders, but also on the drift.

This model is designed to develop instabilities, and numerical simulations
show that, when the drift parameter $\epsilon$ reaches a critical
value $\epsilon_{\rm crit}$, the model does indeed crash, either upwards
or downwards.  It has the interesting feature that, somewhat like phase
transitions in physical materials, or in traffic patterns, the approach
to this transition is characterized by interludes during which the behavior
of the system crosses over to the new (crashing) behavior,
which increase in duration as $\epsilon$ approaches $\epsilon_{\rm crit}$.
The phase transition to this new behavior occurs because
the drift velocity has grown larger than the diffusive velocity,
and begins to dominate the market's movements.  Dimensional analysis
suggests that ${\epsilon}_{\rm crit} \sim J^{1/2}$, which makes sense --
crashes are more common in thin markets than in liquid ones.

The instability in this model is analogous to a particle rolling off of a
sphere, in which the crash may take an arbitrarily large time to develop.
Although any positive bias eventually leads to a crash, we regard as
unstable only those models which lead to a crash within the prediction
horizon of our model, which can be no longer than a few hours, and
certainly cannot extend past the market close.

There are several interesting problems to attack.  The first is
the new, crashing behavior itself -- we would like to understand
how a market's behavior is altered during such a crash.  The simplest
way to do this is to measure the liquidity scaling laws during the
crash, and see how they are changed.  Dimensional analysis suggests
that they should be modified, because the drift $\epsilon$ itself is a new
dimension-ful parameter whose presence now allows a much wider variety of
behavior.  In such a phase, market order traders would be expected to
play a large role and thus should be included in any simulation, as in
the Two-Fluid Model (see the previous section).

An even more interesting problem would be how the market's behavior
is modified in the prelude to a crash.  Again, liquidity scaling
laws would be the simplest quantity to measure in such a case, and
again dimensional analysis suggests that they should be modified
appreciably in the prelude to a crash, within the framework of
our model.  It is plausible that momentum-trading is what becomes
important at such moments, and were such a conjecture born out
by empirical tests in the market, its utility would be obvious.

In addition to the liquidity scaling laws, there is the behavior of
the drift $\epsilon$ itself.  In this simple model, drift feeds on itself.
The parameter $\epsilon$ sits inside a meta-stable basin for a while, but
is buffeted back and forth by chance movements in the market, until it
is knocked out of its basin, and into the unstable region near
${\epsilon}_{\rm crit}$.  Once here, it gathers momentum under its own
steam.  Its trajectory in the unstable region would be an interesting
quantity to calculate.

The bias model describes the onset of crashes, but not their eventual
petering out, because in this model traders remain in the drift-dominated
crashing mode until there are no traders left.  An interesting extension
of this model would be one in which drift is somehow dynamically restored
to small values, allowing the market to return to the quiescent state.  A
model of this kind would enable us to estimate the magnitude and duration
of crashes, because with this new feature they are now only of finite
size.

{\bf Market Makers}

The notion of diffusing traders, who change their trade prices often
before trading, may be considered also in the context of a market with
a single monopolistic market maker, or even several competing market
makers.  This problems has been studied in great detail, from the
point of view of inventory\cite{Garman, Stoll, HoStoll, OharaOldfield,
CMSW78, AmihudMendelson,Hamilton76}, strategic traders\cite{strategy}, and 
in the
presence of information imbalance\cite{infoimb}.

We do not intend to review all this work here, we merely suggest that
in the types of markets alluded to in the introduction, dominated by
day-trading traders, the modeling of the order arrival process might
be modified.  More specifically we propose to refine the modeling of
the order arrival process by using the information contained in the
book of limit orders, together with the diffusion proposal.  To this
end, we demonstrate this calculation in a toy model, which omits all
the other effects mentioned above, inventory, trader strategy, and
information imbalance.

We imagine a market maker who is allowed to see every limit order
available in the market.  As a first approximation, we assume that only
limit order traders are present in the market.  The market maker may set
his bid and offer anywhere he likes.  He desires to maximize his profit,
while minimizing the possibility of any open position.  The market maker
will set his constant bid and offer optimally for a time horizon $T_H$.
We may define, during this horizon, the expected number of shares he will
sell, as
\bq
\Phi_S = \int_{t_0}^{T_H} dt \int_{L_B}^\infty dx' 
G'(L_B,x';t,t_0)\rho_S(x',t_0)
\eq
where $G(x,y;t,t')$ is the Green's function for the diffusion operator,
which satisfies Dirichlet  $G = 0$ boundary conditions at $x = L$.
Similarly the expected number of shares he will buy is
$$
\Phi_B = \int_{t_0}^{T_H} dt \int^{L_S}_0 dx' 
G'(L_S,x';t,t_0)\rho_B(x',t_0)
$$
In terms of this quantities, his expected profit is
$$
P_N = ( L_S - L_B ) {\rm min}( \Phi_B, \Phi_S )
$$
and his expected open position is
$$
P_O = |\Phi_B - \Phi_S|
$$
Thus, we would like to maximize $P_N$ while enforcing $P_O=0$.
We may formulate this as a minimization problem using the technique of
Lagrange multipliers, as  $L_S, L_B, \lambda$ such that
$$
F = P_N - \lambda P_O
$$
is minimized
However, it is more simply reformulated as $L_S, L_B, \lambda$ such that
$$
\tilde F = L_B \Phi_S - L_S \Phi_B - \lambda(\Phi_B - \Phi_S)
$$
is minimized, because the functionals $F$ and $\tilde F$ have the same 
minima.
Minimization of $\tilde F$ leads to the equations
\bq
\Phi_S + (L_B + \lambda) { \partial \over \partial L_B } \Phi_S = 0
\eq
\bq
\Phi_B + (L_S + \lambda) { \partial \over \partial L_S } \Phi_B = 0
\eq
\bq
\Phi_B = \Phi_S
\eq

These equations require numerical solution.

\newsection{Towards Experimental Tests of These Conjectures.}

At the heart of this paper is a proposal that experiments be carried out,
testing the results obtained above and in future publications, regarding
scaling laws in market microstructure, for the markets in which we are
interested.  We describe here what is intended, and will report on work
in this direction in a future publication\cite{ftrexptl}.

The experimental tests of the above results may be done by testing
against real market data, over the short time scales necessary to
determine the elementary statistical fields $B$, $O$, $\chi$, and their
correlations.

Having stated several interesting conjectures in the previous
sections, we desire to test them experimentally.  This appears to pose
some difficulty -- it is not immediate that any of these experiments
can be done.

The reason is that the expectations in the above model are ensemble
averages.  That is, they are averages in which the same market is
started in the same state many times over, and allowed to probabilistically
evolve.  The average is then taken over all the different possible
``worlds'' that these different probabilistic paths may take.  However,
in the real world, there is only one path.  We do not have the
ability to sample many different worlds.  The only averaging
we can do is time-averaging, which in principle is quite a different
thing.

However, we can show by a simple argument that, for certain well-behaved
quantities, they are in fact, the same thing.  This occurs because,
for these quantities, the system is ``forgetful'', so that the
system, after a while, has reset itself, and after the forgetting
time $T$, is actually the same as another version of the initial
world.  In other words, after $N$ forgetting times $T$, we have actually
averaged over $N$ ensemble members.

We show this by considering the ensemble fluctuations of the time average 
of
an operator, and show that they go to zero, under certain conditions.
This means that the operator is diagonal on any state, i.e. a constant
times the identity matrix, and in that case it's ensemble
average must be equal to the constant.  By time translation invariance,
this constant is also equal to the expectation of the (un-time-averaged)
operator.

Notice that we require time translation invariance, here.  If the
state is not at least approximately in a steady state, then the
theorem is just plain false.  However, no one would expect to
get a good experimental measurement, either, of any of the elementary
statistical fields, if the state were to sharply change it's character
during the measurement.  Thus a steady state, or a quasi-steady-state,
is a necessary precondition for both measurement of a time-averaged
quantity and comparison
of that quantity to the ensemble averages we propose to calculate.

The time average of an operator
${\it O}(t)$ is $(1/T)\int_t^{t + T}  {\it O}(\tau)d\tau$.  Notice that it 
is still
an operator, and therefore contains information about all the
different possible worlds defined by the different paths the world
might take.  But as $T \longrightarrow \infty$ this operator often becomes
a scalar times the identity matrix.  To see this, we compute it's
fluctuations in some time translation invariant state.
\bq
{\rm Fluc}( \bar O ) = < \bigl( {1 \over T} \int_t^{t+T} d\tau O(\tau)  -
< {1 \over T} \int_t^{t+T} d\tau O(\tau)> \bigr)^2>
\eq
\bq
= < \int_t^{t + T} d\tau \int_t^{t + T} d\tau' O(\tau) O(\tau') > -
< \int_t^{t + T} d\tau O(\tau) > ^2
\eq
\bq
= \int_t^{t + T} d\tau \int_t^{t + T} d\tau' {\rm Corr}( O(\tau) O(\tau') )
\eq

Thus we see that, if the correlation of $O$ at different times falls off
with $T$, the time average will also fall off to zero, as $T$ gets large.
Hence we arrive at the condition that the only quantities we may compare
with experiment are those whose time-time correlation functions decays
with time $T$.  Fortunately, this is true of most such quantities --
indeed their correlations usually decay exponentially.

Using this result, we may propose a method of measuring the mooted
scaling laws.  We use the bid-offer spread as an example to illustrate
the method.  We record the bid-offer spread, and traded volume at
every instant of time (with a time spacing small enough to catch every
movement of bid and offer prices).  We find an averaging time
sufficient to satisfy the time-averaging relation above, and compute
from this recorded data a series of time averages of spread versus
deal rate.  We then make a scatter plot of $\log S$ versus $\log J$,
and compute a best fit line to this scatter plot -- the prediction of
the minimal model is that this slope should be -1/2.

The bid-offer spread in dealer markets, as opposed to the double
auction of this paper, has been studied empirically by
\cite{Demsetz,Tinic, TinicWest74, TinicWest72, Stoll_Exp, AmihudMendelson,
BenstonHagerman, NewtonQuandt, Hamilton76, Hamilton78, BranchFreed}.
These studies were motivated by debates on market design.  For
example, the restrictions on trading of the registered traders in the
Toronto Stock Exchange (TSE), compared against the affirmative
obligations for members of the New York Stock Exchange (NYSE), and of
the more independent dealers of the OTC was studied in
\cite{TinicWest74}.  The behavior of the spreads of the specialists on
the NASDAQ vs. the dealers in the OTC was studied by
\cite{Hamilton78}, the spreads on the AMEX were compared with those of
the NYSE by \cite{BranchFreed}, and the issue of competitive dealing
on the NYSE was studied by \cite{TinicWest72, Hamilton76}.  This field
of study was in fact originated by Demsetz, who studied dealer spreads
on the NYSE\cite{Demsetz}.  Each of these studies consisted of a
linear regression applied to a large number of markets in individual
stocks, and showed (among other things) that spreads have a negative
gradient with deal rate, as we expect in our markets.  In the case of
\cite{Demsetz,TinicWest72,Hamilton76}, this was used to demonstrated
the industry-wide, but not firm-wide, scale economies that exist in
the market for dealer serveices, and made the case against natural
monopolies, in favor of competitive dealer services.

\newsection{Conclusion}

In this paper we have introduced into the study of market
microstructure a series of models for markets made up of professional
traders.  These traders scrap for every basis point, and trade
principally through limit orders, which they change frequently in
response to the changing position of their book, their various
perceptions of the direction of the market, and movements of different
but related markets.  This type of trader typically trades on behalf
of an institution, such as a bank in, e.g. the interdealer broker
markets.  We modeled their changing limit orders as random walks which
terminate when they collide with a price of the opposite kind, a buyer
with a seller.  This kind of dynamics, known as diffusion-annihilation
dynamics, has been well-studied in the context of physics and
chemistry.

We used this modeling framework to study the scaling of
quantities related to market microstructure.  Our expectation is that
the purely statistical effects of having large numbers of traders in a
market will be the dominant effect for these quantities, and we do not
entertain the more difficult task of seeking the more subtle effects of
trader behavior.  Scaling laws quantify the coarsest features of the 
markets
behavior, and do not require a detailed model.  Hence we focused our
efforts on these scaling laws.

Our scaling laws were evaluated as expectations in a measure.  This
measure was defined as the steady state solution of our dynamics, under
the reasoning that a quiescent market, one not undergoing a large sudden
movement, but only trading at an approximately constant rate, was like
a steady state market.  Furthermore, it is only in a steady state
market that these scaling laws would be well-defined and meaningful.

We began by studying the scaling laws for various proxies for liquidity,
such as the bid-offer spread ${\rm Spr}$, time to midmarket trade $\tau$,
trader density near the best bid/offer $\Xi_{(\cdot)}$, and midmarket
variance $\xi$.  We found ${\rm Spr} \sim J^{-1/2}$, $\tau \sim J^{-1}$,
$\Xi_{(\cdot)} \sim J^{1/2}$, and $\xi \sim J^{-1/2}$.  We then went on
to study more complex correlations and conditional expectations as a
function of $J$, whose qualitative behavior had a ready intuition.  These
included the correlation of time changes in best bid with time changes in
best offer $C_{\Delta_t B \; \Delta_t O}$, the density at the best
bid/offer conditioned on fluctuation in best bid/offer
$\Xi_{(\cdot)}(\xi_{(\cdot)})$, and the time change in the bid/offer spread
conditioned on the bid/offer spread $\Delta_t {\rm Spr}({\rm Spr})$. In
the latter two cases we found the scaling forms
$\Xi_{(\cdot)}(\xi_{(\cdot)}) \sim \sqrt{ J\over D} f( {D \over J
\xi^2_{(\cdot)}})$, and $\Delta_t {\rm Spr}( Spr = s_0 ) \sim \sqrt{ D
\over J }  f( {D \over J s_0^2})$.

We also examined response functions to density imbalance, estimated
the size a tender offer must be to attract a certain number of buyers,
and the expected profits of a specialist.  Although we have, by
dimensional analysis, extracted some information from our model, a
full treatment of these scaling laws requires numerical simulation,
the results of which shall appear in a forthcoming publication.

Some of our scaling laws were amenable to treatment by an
approximation developed in \cite{cardy}.
We calculated the variance of the midmarket $w$, and found a
logarithmic correction factor to our dimension-analytic result. The
method also enabled us to calculate the time correlation function.

The above scaling laws and scaling forms were studied first in a model
with as few parameters as possible, to isolate the purely statistical
effects.  We then introduced to this model, the `` minimal model'',
a series of other features intended to explore its response under more
realistic market conditions, such as the influence of random events.
We considered the influence of randomness in the trader's drift, as
due to the presence of shocks from news items, or the influence of
related markets.  Then we considered the influence of a randomly
fluctuating rate of entry of new traders into the market, centered
around some average.  Finally, we considered the diffusion constant
itself to vary randomly.  We assumed that the fluctuations from the
noise were small, so as to generate small corrections to the minimal
model.

In addition, we introduced to the minimal model the influence of ``market
order traders'', those who do not register a visible bid or offer in the
market before trading.  This led us to a two-fluid model, in which the
dynamics is very similar to that of the minimal model, but the invisible
traders cannot react with one another, and the bid-offer spread is
measured only between the visible traders.  We proposed this model for
the purpose of investigating scaling laws during a crash, as though this
might have its own sort of steady state behavior.  We may cause it to 
crash by
introducing a large flux of sellers, and we will see that the bid-offer
spread widens out in the expected manner, and it would be interesting to
use this model to investigate crossover behavior between crashing and
non-crashing regimes.

We then added in a drift term to the trader's random walk, whose novel
feature is that it changes with the changes in the trade price.  This
feature is intended to model momentum-trading, or herd-like behavior
in traders, and requires that we consider slightly longer time scales,
over which these influences may exert themselves.  It causes the model
to develop crash-like instabilities, stemming from a transition from
diffusion dominated dynamics to drift dominated dynamics.  It is
interesting to study the scaling laws in the prelude to a crash, to
observe how they may signal a market's instability.  We propose a
series of calculations of this kind.  They must, however be done
numerically, and we postpone these to a future publication
\cite{ftrnumerical}.

Finally, we examined the possibility of treating the more traditional
dealer problem in the context of interdealer broker markets, using the
proposal of diffusing bids and offers.  We found that it allowed us to
use the additional information contained in the limit order book to make
a more refined model for the order arrival process, and we illustrated
this with a toy model in which the dealer's optimization problem for the
bid-offer spread was defined.

In future publications, we shall do computer simulations to
investigate the more detailed predictions of these models, and compare
them directly with live market data\cite{ftrexptl}.

There are several other directions in which research might proceed.  One
possibility is to observe a market's reaction in response to a shock.  In
our minimal model, this would correspond to the system being suddenly
moved out of its ground state, and partially populating the first few
excited states.  Because the dynamics of this model are of diffusion
type, these excitations will fade away exponentially, with characteristic
decay rates given by the ``energies'' of the excited states.  By
computing these energies, it is possible to predict the decay rate of
market excitations.  In a future publication we shall attempt to
calculate these decay rates, and again compare them with market data.

Another possibility for future research is to refine the minimal model.  We
may add refinements in several ways.  One way might be, following
Garman \cite{Garman}, to introduce a rate $\lambda$ at which traders at the
point $x$ enter and leave the market, proportional to their population
at $x$.  Another way to refine the model would be to introduce interactions
between the buyers and sellers.  The most obvious kind are those which
allow traders to react to what they see on the trading screen.  We will
discuss in a future publication the ``competitive market''.  This is a
market in which traders are anxious to do deals quickly, and as such
desire to be the best bid or offer.  Bidders are therefore attracted to the
best bid, and offers attracted to the best offer, so that they might do
the deal first. This would most likely tighten market spreads, and change
the scaling law exponents.

Finally, it is interesting to note that the minimal model bears many
similarities to members of a class of exactly soluble models that have
been elucidated by work over the past two decades\cite{BetheAnsatz}.
We will explore in a future publication the application of this method
to the minimal model.

\newsection{ Appendix}

{\it The form of the time evolution operator}

We introduce an operator, $U_{\rm ann}$, the "annihilation operator",
which takes any state which has sellers and buyers at the same point
in price space, and returns the same state with any overlapping buyers
and sellers deleted pairwise.  If in the state no buyers and sellers
overlap, the annihilation operator does not change the state.

We also introduce diffusion separately for the buyers and sellers
$D_B(t,t')$, $D_S(t,t')$.  These are the usual diffusion operators.
When acting on any initial state, the result is a solution to the
diffusion equation.  However, we cannot allow these diffusion
operators to act for very long periods, because at any moment they
might cause a seller to jump onto a buyer, or vice versa.
Therefore, our diffusion-annihilation operator must diffuse for a very
short time, then annihilate, then diffuse some more, annihilate
again, etc.  We alternate diffusing buyers and sellers, and always
insert an annihilation operator in between.  Thus, our evolution
operator must take the form
\bq
Ev(t,t') = \lim_{\delta t \longrightarrow 0} \prod_{n=0}^{(t'-t)/\delta t}
U_{\rm ann} D_S(t + n \delta t,t + (n+1) \delta t)
U_{\rm ann} D_B(t + n \delta t,t + (n+1) \delta t)
\label{evform}
\eq

Sometimes we will additionally insert into our state a buyer or seller 
that does
not diffuse, but moves according to a prescribed motion.  Then this trader 
must
be regarded as a different species, with its own separate evolution, which
we mix into the evolution operator.  Let the operator defining this 
prescribed
motion be $D^{(\cdot)}_p(t',t)$, where $(\cdot)$ may be either $S$ or $B$.
We imagine that the annihilation operator has been appropriately adjusted 
to
take into account annihilations between buyers and sellers of all 
species.  Then
$Ev$ becomes
\bq
Ev(t,t') = \lim_{\delta t \longrightarrow 0} \prod_{n=0}^{(t'-t)/\delta t}
U_{\rm ann} D^S_p(t + n \delta t,t + (n+1) \delta t)
U_{\rm ann} D^B_p(t + n \delta t,t + (n+1) \delta t)
\nonumber
\eq
\bq
U_{\rm ann} D_S(t + n \delta t,t + (n+1) \delta t)
U_{\rm ann} D_B(t + n \delta t,t + (n+1) \delta t)
\eq
Obviously, we may introduce as many species as required, and the above
provides a template for how we must adjust the evolution operators to
properly evolve this more complicated system.

{\it The Form of the Diffusion Operators $D_B$, $D_S$}

Diffusion operators occur commonly in statistical field theory, and
their form is well known, see, for example, \cite{Creutz}.  We give
the standard result here, for completeness.
\bq
D_B(t,t') = \exp{ \bigl( H^{(B)}_{\rm Diff}(t-t')\bigr) } ~~~~
H^{(B)}_{\rm Diff} = \sum_x \Psi_B^\dagger(x+1)\Psi_B(x) + 
\Psi_B^\dagger(x-1)\Psi_B(x)
\eq
\bq
D_S(t,t') = \exp{ \bigl( H^{(S)}_{\rm Diff}(t-t')\bigr) } ~~~~
H^{(S)}_{\rm Diff} = \sum_x \Psi_S^\dagger(x+1)\Psi_S(x) + 
\Psi_S^\dagger(x-1)\Psi_S(x)
\eq

{\it The Form of the Annihilation Operator $U_{\rm ann}$ }

The annihilation operator is an operator that, at every point,
eliminates pairs of buyers and sellers that sit on top of one
another.  After the action of $U_{\rm ann}$ every point contains
either buyers only or sellers only.

$U_{\rm ann}$ can be written as a product of operators $U_{\rm
ann}(x)$ at each point $U_{\rm ann} = \prod_x U_{\rm ann}(x)$, whose
job is to annihilate pairs at the point $x$.  A simple operator that
accomplishes this is
\bq
U_{\rm ann}(x) =
(\Psi^{(B)}_x \Psi^{(S)}_x)^{\min( N^{(B)}_x, N^{(S)}_x ) }
\eq
however, this is not particularly analytically tractable.  We may
develop another form by considering the master equation for chemical
reactions at a point $x$, which occur at rate $\lambda$.  Let
$P(n_b,n_s)$ be the probability of having $n_b$ buyers and $n_s$
sellers at point $x$.  Consider a price space with only a single
point.  Then the reaction probability per unit time is proportional to
the number of particles present $\sim \lambda n_b n_s P(n_b, n_s)$, so
that
\bq
{\partial P(n_b,n_s) \over \partial t} = \lambda \bigl( (n_b + 1)(n_s +
1)P(n_b+1,n_s+1)- n_b n_s P(n_b,n_s) \bigr)
\eq
For a price space with many points, and a state described by a
probability functional $P({n_b},{n_s})$ which is a functional of
integer sequences ${n_b}$, ${n_s}$ on the price space ${\cal P}$, this
is easily generalized to
\bq
{\partial P({n_b},{n_s}) \over \partial t} =
\lambda \sum_x \bigl( (n_b(x) + 1)(n_s(x) + 1)P({n_b+1},{n_s+1})-
n_b(x) n_s(x) P({n_b},{n_s}) \bigr)
\eq
This may be expressed in terms of the action of the field operators
$\Psi_B(x)$,$\Psi_S(x)$, $\Psi_B^\dagger(x)$,$\Psi_S^\dagger(x)$, as
\bq
{\partial P({n_b},{n_s}) \over \partial t} = H_{\rm ann} P({n_b},{n_s})
\eq
where $H_{\rm ann}$ is
\bq
H_{\rm ann} = \sum_x H_{\rm ann}(x) ~~~~~
H_{\rm ann}(x) = \lambda (\Psi_B^\dagger(x)\Psi_S^\dagger(x) - 
1)\Psi_B(x)\Psi_S(x)
\eq
so that $U_{\rm ann}(x) = \exp{\bigl( H_{\rm ann}(x) \Delta t
\bigr)}$, where $\Delta t$ is the time interval over which this
operator is allowed to act.

In general, an annihilation operator of this form does not completely
annihilate the particles we want annihilated, but the population of
such particles decays exponentially, at the rate $\lambda$.  Because
of this, sellers can sometimes overlap buyers, a feature which is
unrealistic.  However, by taking the rate $\lambda \Delta t$ to
be very large, this effect can be made arbitrarily small.

The reader should note that this annihilation operator annihilates
buyers of a specific type with sellers of a specific type.  When there
is more than one type, and consequently more than one reaction pair,

then we must introduce an annihilation operator for each pair, and the
full annihilation operator is the product of all of them.

{\it Fields with Memory: The Instantaneous Deal Rate Operator, and the 
Last Trade Field}

On a trading screen, one often sees the price at which the most recent
trade occured, regardless of how long ago it may have taken place.  We
may also infer from a trading screen the rate at which trades occur,
on average, the ``deal rate''.  We demonstrate how these and related
quantities, that require some memory in the system, may be calculated
within this kind of model.  It will prove convenient to compute these
in terms of an operator $N_\zeta(x,t)$, which, when acting on a
pure state at time $t$ increases by $1$ at point $x$ if a trade has
occurred at time $t$ and price $x$, and zero otherwise.  The
instantaneous deal ra
te $J(t)$ is then $J(t) = \sum_x'
N_\zeta(x',t)$, and the cumulative deal rate $J_{\rm cum}(t)$ is
$J_{\rm cum}(t) = \int^t dt' \sum_x' N_\zeta(x',t')$.  $J_{\rm
cum}(t)$ is a nondecreasing function of $t$, and as such we may define
an inverse mapping $J^{-1}_{\rm cum}(t)$, which associates to a total
deal volume the {\it first} time it was achieved.

The last trade field $\chi(x,t)$ is constructed as follows.  Consider
$\tilde \chi(t) = \sum_x' x' (d/dt) N_\zeta(x',t)$.  If trades occur
at one and only one trade price at every time slice $t$ (as they must),
then $\tilde \chi(t)$ is equal to, at time t, $x$ if a trade
occured at time $t$, and 0 otherwise.  This is unsatisfactory for
the last trade field, because we want to know the price
at which the last trade occurred, regardless of how long ago it occurred.
$\tilde \chi(t)$ may be thought of as the ``non-persistent'' last trade
field.  The ``persistent'' trade field $\chi(t)$, the one we are
after, is then related to the nonpersistent trade field $\tilde \chi(t)$
by $\chi(t) = \tilde \chi( J^{-1}_{\rm cum}( J_{\rm cum}(t)))$.

The operator $N_\zeta$ introduced above is the number operator for
a new field $\zeta$, with destruction $\zeta_x$ and creation operator
$\zeta^\dagger_x$, $N_\zeta(x) = \zeta^\dagger_x \zeta_x/ \delta S$.  In
order that $N_\zeta(x)$ satisfy what is required of it, we must
insert a $\zeta$ where a trade occurs.  To do this, we modify the
annihilation operator $U_{\rm ann}$.  The pairwise annihilation
factors each become
\bq
H_{\rm ann}(x) = \lambda (\Psi_B^\dagger(x)\Psi_S^\dagger(x) - 1)
\zeta^\dagger(x)\Psi_B(x)\Psi_S(x)
\eq

Note that the field $\zeta$ differs from the others we have
introduced, because it does not represent a trader.   It is a
mathematical device that records trading information in the position
of the field $\zeta$.

{\bf Acknowledgements.}

It is a pleasure to thank Judith Bender, Jean-Philippe Bouchaud,
Frank Brown,John Cardy, Marc Devorsetz,  Amy Eliezer, Fabian Essler,  Andrew Graham,
  Ron Kantor, James Lee, Naresh
Malhotra, Peter Orland,  Zoltan Racz, Magnus Richardson, Sven Sandow,
 Fred Schmidt, David Sherrington, Robin Stinchcombe and David Vines
for useful discussions  and interest in  this project. We would like
to thank John Cardy,  Zoltan Racz and   Magnus Richardson  for
 elucidating  the theory of  diffusion-controlled annihilation.
  We are indebted to David Sherrington for  
invaluable help during the preparation of the final version of this paper.

\end{document}